\documentclass[sigconf]{acmart}
\AtBeginDocument{%
  \providecommand\BibTeX{{%
    \normalfont B\kern-0.5em{\scshape i\kern-0.25em b}\kern-0.8em\TeX}}}

\copyrightyear{2021}
\acmYear{2021}
\setcopyright{acmlicensed}
\acmConference[SIGIR '21] {Proceedings of the 44th International ACM SIGIR Conference on Research and Development in Information Retrieval}{July 11--15, 2021}{Virtual Event, Canada.}
\acmBooktitle{Proceedings of the 44th International ACM SIGIR Conference on Research and Development in Information Retrieval (SIGIR '21), July 11--15, 2021, Virtual Event, Canada}
\acmPrice{15.00}
\acmISBN{978-1-4503-8037-9/21/07}
\acmDOI{10.1145/3404835.3462920}

\usepackage[ruled,vlined,linesnumbered]{algorithm2e} 
\usepackage{subfigure} 
\usepackage{amsmath} 

\setcopyright{none}
\renewcommand\footnotetextcopyrightpermission[1]{} 
\newcommand{\agent}{agent}
\newcommand{\arm}{arm}
\newcommand{\conversation}{conversation}
\newcommand{\myalg}{RelativeConUCB}

\newcommand{\suparm}{key-term}
\newcommand{\firstuppersuparm}{Key-term} 
\newcommand{\updatepos}{Pos}
\newcommand{\updateposneg}{Pos\&Neg}
\newcommand{\updatediff}{Difference}
\newcommand{\user}{user}

\newcommand{\armsymbol}{a}
\newcommand{\suparmsymbol}{k}
\newcommand{\usersymbol}{u}
\newcommand{\bernoulli}{{\rm Bernoulli}}

\newcommand{\gaussian}{\mathcal{N}}
\newcommand{\defeq}{\stackrel{\bigtriangleup}{=}}

\newcommand{\bias}{b}
\newcommand{\feedback}{\tilde{r}}
\newcommand{\individual}{\beta}
\newcommand{\noise}{\epsilon}
\newcommand{\reward}{r}
\newcommand{\scale}{c}

\newcommand{\armvec}{\boldsymbol{x}}
\newcommand{\armlabelvec}{\boldsymbol{b}}
\newcommand{\suparmvec}{\tilde{\boldsymbol{x}}}
\newcommand{\suparmpseudovec}{\dot{\boldsymbol{x}}}
\newcommand{\suparmlabelvec}{\tilde{\boldsymbol{b}}}
\newcommand{\uservec}{\boldsymbol{\theta}}
\newcommand{\userarmvec}{\boldsymbol{\theta}}
\newcommand{\usersuparmvec}{\tilde{\boldsymbol{\theta}}}

\newcommand{\armmat}{\mathbf{X}}
\newcommand{\armdatamat}{\mathbf{M}}

\newcommand{\suparmdatamat}{\tilde{\mathbf{M}}}
\newcommand{\weightmat}{\mathbf{W}}

\newcommand{\armset}{\mathcal{A}}
\newcommand{\R}{\mathbb{R}}

\newcommand{\suparmset}{\mathcal{K}}
\newcommand{\userset}{\mathcal{U}}
\newcommand{\observeset}{\mathcal{D}}

\begin{document}
\fancyhead{}

\title{Comparison-based Conversational Recommender System with Relative Bandit Feedback}

\author{Zhihui Xie}
\affiliation{
  \institution{Shanghai Jiao Tong University}
  \city{Shanghai}
  \country{China}
}
\email{fffffarmer@sjtu.edu.cn}

\author{Tong Yu}
\affiliation{
  \institution{Carnegie Mellon University}
  \city{Pittsburgh}
  \state{Pennsylvania}
  \country{USA}
}
\email{worktongyu@gmail.com}

\author{Canzhe Zhao}
\authornote{The work is done while the student is an intern at Shanghai Jiao Tong University.}
\affiliation{
  \institution{Shandong University}
  \city{Jinan}
  \state{Shandong}
  \country{China}
}
\email{zcz@mail.sdu.edu.cn}

\author{Shuai Li}
\authornote{Corresponding author.}
\affiliation{
  \institution{Shanghai Jiao Tong University}
  \city{Shanghai}
  \country{China}
}
\email{shuaili8@sjtu.edu.cn}


\begin{abstract}
With the recent advances of conversational recommendations, the recommender system is able to actively and dynamically elicit user preference via conversational interactions. To achieve this, the system periodically queries users' preference on attributes and collects their feedback. However, most existing conversational recommender systems only enable the user to provide absolute feedback to the attributes. In practice, the absolute feedback is usually limited, as the users tend to provide biased feedback when expressing the preference. Instead, the user is often more inclined to express comparative preferences, since user preferences are inherently relative. To enable users to provide comparative preferences during conversational interactions, we propose a novel comparison-based conversational recommender system. The relative feedback, though more practical, is not easy to be incorporated since its feedback scale is always mismatched with users' absolute preferences. With effectively collecting and understanding the relative feedback from an interactive manner, we further propose a new bandit algorithm, which we call \textit{\myalg}. The experiments on both synthetic and real-world datasets validate the advantage of our proposed method, compared to the existing bandit algorithms in the conversational recommender systems.

\end{abstract}



\maketitle


\section{Introduction}


\begin{figure}[t]
    \centering
    \includegraphics[width=\linewidth]{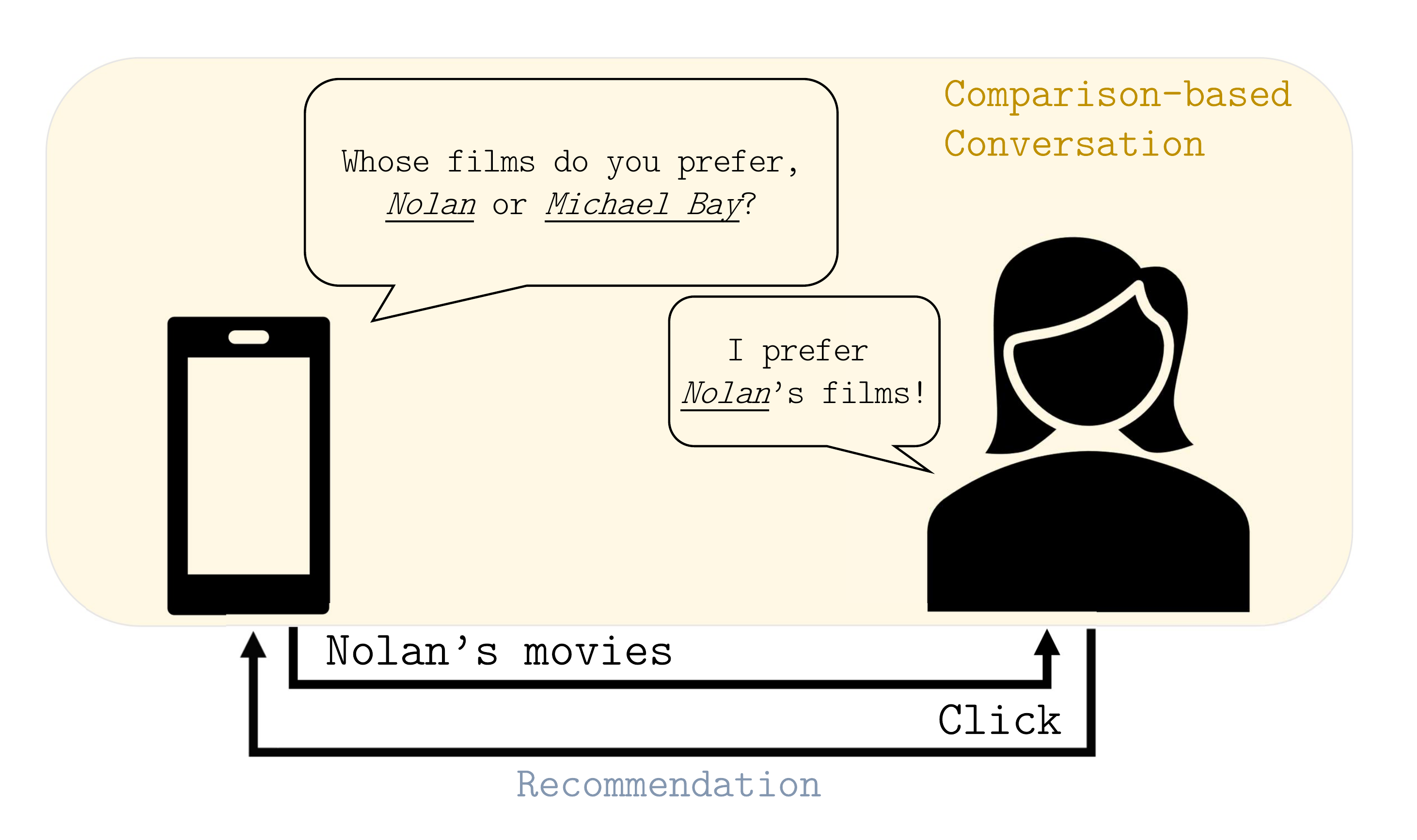}
    \caption{An illustrative example of a comparison-based conversational recommender system.}
    \label{fig:illustration}
    \Description[An illustration for the proposed framework]{An illustration for the proposed comparison-based conversational recommender system.}
\end{figure}

Recommender systems have received enormous attention in this age of big data. In many scenarios, for a recommender system to interact adaptively and optimally with users, it must learn effectively from user feedback.
However, for cold-start users with little historical data, it is very challenging to recommend optimally in an online manner.
Moreover, the data collected from users during recommendation can be extremely skewed, as users only express their preference on the recommended item.
To better elicit users' preference, the idea of conversational recommender system (CRS) was proposed \cite{christakopoulou2016towards}.
In addition to collecting responses (e.g., click behaviors) on recommended items, the system also periodically conducts conversations with the user to better elicit preference.
This allows the system to provide personalized recommendations based on the responses.

Despite the success of CRSs in recent years, we argue that current conversation mechanism is still limited.
Most existing CRSs only allow the user to provide absolute feedback, from which it can be hard to collect useful and accurate information in practice \cite{joachims2017accurately, radlinski2008does, chapelle2012large, yue2009interactively}.
For example, in a movie recommendation scenario, the {\agent} may query about the user's preference on a particular category: ``\textit{Do you like Nolan's movies?}''.
As a big fan of Nolan, the user would answer positively.
But for a similar question ``\textit{What about Michael Bay's movies?}'', it could be difficult to answer as the user holds a very mixed attitude towards the director's works.
The love felt for the series of Transformers would eventually induce another positive answer.
The contradictory feedback misleads the {\agent} to believe that the user admires both directors equally.
To address this issue, we consider asking relative questions to elicit the user's preference.
We argue that relative feedback can be more accurate, since user preferences are inherently relative.

Motivated by the above considerations, we propose to build a comparison-based conversational recommender system.
To the best of our knowledge, few works have reached this topic.
The system follows the conventional CRS scenario, recommending items sequentially to the user based on contextual information.
Compared to existing CRSs, the proposed framework differs on how to conduct conversations.
The comparison-based conversational recommender asks relative questions over {\suparm}s, and then receives relative feedback inferring the user's comparative preference.
Similar to that in \cite{zhang2020conversational}, a {\suparm} is a category-like concept, representing a specific group of items.
Movies directed by Nolan is an example of a {\suparm} in the mentioned scenario.

In spite of the mentioned superiority of relative feedback,
it could be difficult to incorporate relative feedback due to the mismatch between comparative signals and the underlying user preferences.
With effectively collecting and understanding the relative feedback from an interactive manner, we further propose a new bandit algorithm, which we call \textit{\myalg}.
{\myalg} formulates conversation and recommendation as a two-stage learning process.
When conducting conversations, {\myalg} selects the pair of {\suparm}s which maximizes error reduction, and queries the user's comparative preference over it.
To incorporate relative feedback, we develop three different mechanisms.
The experiments show that our algorithm performs competitively well compared to existing bandit algorithms in the conversational recommender systems.

In summary, we make three major contributions:
\begin{itemize}
    \item We present a novel comparison-based CRS, which allows users to provide relative feedback to the comparisons of {\suparm}s.
    This extends the framework of existing CRSs.
    \item We propose a new bandit algorithm {\myalg} to build a comparison-based CRS.
    \item Experiments on both synthetic and real-world datasets clearly show the advantages of our method.
\end{itemize}

The rest of the paper is organized as follows.
We specify the considered comparison-based conversational recommender system in Section \ref{sec:problem_formulation}.
In Section \ref{sec:algorithm} we introduce the {\myalg} algorithm in details.
The experimental setup and results for evaluation are presented in Section \ref{sec:experiments}.
Finally, we discuss some related works in Section \ref{sec:related_work} and then conclude our work in Section \ref{sec:conclusion}.
\section{System Overview}\label{sec:problem_formulation}
In this section we will formulate how our comparison-based conversational recommender system works.
We start by introducing the conventional contextual bandit scenario, and then extend the system, equipping it with the power to conduct comparison-based conversations.

\subsection{Contextual Bandit}
Contextual bandits' objective is to optimize the expected cumulative rewards in a long run.
To achieve this, they need to acquire enough information about {\arm}s while simultaneously utilizing the available knowledge to choose the best {\arm}.

Suppose there is a finite set of {\arm}s (e.g., movies in the domain of movie recommendation) denoted by an index set $\armset$ in the system.
The {\agent} will interact with a user with unknown feature $\uservec^* \in \R^d$ consecutively for $T$ rounds.
At each round $t$, the {\agent} only has access to a subset of {\arm}s $\armset_{t} \subseteq \armset$, from which the {\agent} needs to choose an {\arm} $\armsymbol_{t}$ based on its observed contextual vector $\armvec_{\armsymbol_{t}} \in \R^d$ and recommend it to the {\user}.
The {\agent} will then receive a binary reward representing the click behavior:
\begin{equation}\label{eq:reward}
    \reward_{\armsymbol_{t}, t} \sim
    \bernoulli(\armvec_{\armsymbol_{t}}^\top \uservec^*).
\end{equation}

The target of the {\agent} is to maximize the cumulative reward or, equivalently, to minimize the cumulative regret in $T \in \mathbb{N}_{+}$ rounds of interaction:
\begin{equation}\label{eq:regret}
    R \defeq \mathbb{E}\left[\sum_{t=1}^{T} \max _{a \in \mathcal{A}_{t}}\armvec_{a}^\top \uservec^*-\sum_{t=1}^{T} \reward_{\armsymbol_t, t}\right],
\end{equation}
where regret accumulates as sub-optimal arms are selected.

\subsection{Comparison-based CRS} 
A recent work \cite{zhang2020conversational} extends the contextual bandit scenario for conversational recommendation.
It allows the {\agent} to occasionally conduct {\conversation}s with {\user}s, asking the {\user}'s preference over \textit{{\suparm}s}.
Specifically, a {\suparm} can be a category or a common attribute (e.g., the category of comedy movies in movie recommendation).
{\firstuppersuparm}s in the system are denoted by an index set $\suparmset$.
The relationship between {\arm}s and {\suparm}s are defined by a weighted bipartite graph $(\armset, \suparmset, \weightmat)$, where {\suparm} ${\suparmsymbol}$ is associated to {\arm} $\armsymbol$ with weight $\weightmat_{\armsymbol, \suparmsymbol} \geq 0$.
To normalize, we assume $\sum_{\suparmsymbol \in \suparmset} \weightmat_{\armsymbol, \suparmsymbol} = 1$, $\forall \armsymbol \in \armset$.
The contextual vector for {\suparm} $\suparmsymbol$ is then formulated as $\suparmvec_\suparmsymbol = \sum_{\armsymbol \in \armset} \frac{\weightmat_{\armsymbol, \suparmsymbol}}{\sum_{\armsymbol^{\prime} \in \armset} \weightmat_{\armsymbol^{\prime}, \suparmsymbol}} \armvec_{\armsymbol}$.

\subsubsection{Relative Feedback}
In the above conversational recommender system, the {\agent} is only allowed to ask the user absolute questions like ``\textit{Are you interested in Nolan's films?}''
We argue that it is usually unpractical, as the user may have very biased attitudes towards the {\suparm}s, in which case the {\agent} can only collect limited information.
Comparative preferences in the meanwhile are potentially more informative and cheaper to collect.
Therefore, we design the comparison-based conversation scenario.
To conduct conversations, the {\agent} queries relative questions over pairs of {\suparm}s.
For example, as illustrated in Figure \ref{fig:illustration}, the {\agent} may ask ``\textit{Whose films do you prefer, Nolan or Michael Bay?}''.
It then receives the feedback inferring the user's comparative preference.
To simplify, we assume no abstention, making the feedback binary.
Furthermore, we assume the feedback may be corrupted by Gaussian noise.
Suppose the {\user} has an internal valuation towards {\suparm} $\suparmsymbol$:
\begin{equation*}
    \feedback_{\suparmsymbol, t} = \suparmvec_{\suparmsymbol}^\top \uservec^* + \noise_{\suparmsymbol, t},
\end{equation*}
where $\noise_{\suparmsymbol, t} \sim \gaussian(0, \sigma_g^2)$ denotes the random noise. The relative feedback for {\suparm} $\suparmsymbol_1$ and $\suparmsymbol_2$ is:
\begin{equation}\label{eq:relative_feedback}
    \feedback_{\suparmsymbol_1, \suparmsymbol_2, t} = \mathbf{1}\left[\feedback_{\suparmsymbol_1, t} > \feedback_{\suparmsymbol_2, t}\right].
\end{equation}

\begin{algorithm}[t]
\SetKwInOut{Input}{Input}

\Input{Arms $\armset$, {\suparm}s $\suparmset$, weight $\weightmat$, users $\userset$, and conversation frequency function $b(\cdot)$.}

Initialize the numbers of users' interaction $t_\usersymbol = 0, \forall \usersymbol \in \userset$\;

\For{$i = 1, 2, \dots, N$} {
    An arbitrary user $\usersymbol \in \userset$ arrives with $t = t_\usersymbol$ rounds of historical interaction\label{line:framework-forbegin}\;
    Set the conversation budget $q_t$ according to Equation \ref{eq:budget}\;
    Determine the candidate {\arm}s $\armset_t$ and the candidate {\suparm}s $\suparmset_t$ (e.g., by sampling)\;
    \tcc{An agent starts to interact with $\usersymbol$}
    \For{$j = 1, 2, \dots, q_t$\label{line:interactbegin}}{
        Select a pair of {\suparm}s $\suparmsymbol_1, \suparmsymbol_2 \in \suparmset_t$ to query the user preference\;
        Receive the conversational feedback $\feedback_{\suparmsymbol_1, \suparmsymbol_2, t} \in \{0, 1\}$\;
        Update the model parameters\;
    }
    Select an {\arm} $\armsymbol_t \in \armset_t$ to recommend\;
    Receive the reward $\reward_{\armsymbol_t, t} \in \{0, 1\}$\;
    Update the model parameters\label{line:framework-forend}\;
    Update $t_\usersymbol = t_\usersymbol + 1$\;
}
\caption{Comparison-based Conversational Recommender System}
\label{alg:framework}
\end{algorithm}

\subsubsection{Conversation Frequency}
Similarly to \cite{zhang2020conversational}, we model the {\conversation} frequency by a function $b(\cdot)$.
Specifically, if $b(t) - b(t - 1) > 0$, the {\agent} will conduct 
\begin{equation}\label{eq:budget}
    q_t = b(t) - b(t - 1)
\end{equation}
conversations with user at the $t$-th round of interaction.
In such a manner, the {\agent} will conduct $b(t)$ conversations with the user up to round $t$.

\subsubsection{Long-run Interaction}
To simulate the long-run process of interaction with multiple users, we consider a sequence of $N$ iterations described in Algorithm \ref{alg:framework}.
Assume there is an arbitrary distribution $p$ over {\user}s indicating the user frequency.
At each iteration\footnote{It needs to clarify that `iteration' considers interaction with multiple users, while `round' refers to interaction with a specific user.}, an arbitrary user drawn randomly from $p$ will arrive.
The {\agent}, unknowing the user's real feature, will then interact with the user as described in Line \ref{line:interactbegin}-\ref{line:framework-forend} of Algorithm \ref{alg:framework}.
This framework extends the original contextual bandit scenario by considering one particular user may not arrive consecutively.

The overall objective for an agent in the comparison-based CRS is to minimize the cumulative regret induced by all interacted users in a long run:

\begin{equation}\label{eq:overall_cum_regret}
    R(N) \defeq
    \sum_{\usersymbol \in \userset_N} R_{\usersymbol},
\end{equation}
where index set $\userset_N$ denotes the users that have interacted with the {\agent} for at least one round and $R_{\usersymbol}$ denotes the cumulative regret induced by the historical interaction with user $\usersymbol$ defined in Equation \ref{eq:regret}.
\section{Algorithm}\label{sec:algorithm}
To build a comparison-based conversational recommender system, we propose the {\myalg} algorithm.



\begin{algorithm}[t]
\SetKwInOut{Input}{Input}
\SetKwInOut{Init}{Init}

\Input{
1) User $\usersymbol$ with $t = t_\usersymbol$ rounds of historical interaction, parameters $\armdatamat_\usersymbol$, $\armlabelvec_\usersymbol$, $\suparmdatamat_\usersymbol$ and $\suparmlabelvec_\usersymbol$;\newline
2) candidate {\arm}s $\armset_t$, candidate {\suparm}s $\suparmset_t$, weight $\weightmat$, and conversation budget $q_t$;\newline
3) hyper-parameters $\lambda$, $\tilde{\lambda}$, $\sigma$, $\alpha$, and $\tilde{\alpha}$.
}
\BlankLine

\If{$t = 0$} {
    Initialize $\armdatamat_\usersymbol=(1-\lambda) \mathbf{I}, \armlabelvec_\usersymbol=\mathbf{0}$, $\suparmdatamat_\usersymbol=\tilde{\lambda} \mathbf{I}$, $\suparmlabelvec_\usersymbol=\mathbf{0}$\;
}
\For{$j = 1, 2, \dots, q_t$}{
    Select a pair of {\suparm}s $\suparmsymbol_1, \suparmsymbol_2 \in \suparmset_t$ according to Section \ref{sec:pos-posneg} or Section \ref{sec:difference}\;\label{line:suparm_selection}
    Receive the conversational feedback $\feedback_{\suparmsymbol_1, \suparmsymbol_2, t}$ according to Equation \ref{eq:relative_feedback}\;
    Update $\suparmdatamat_\usersymbol$ and $\suparmlabelvec_\usersymbol$ according to Algorithm \ref{alg:suparmupdate}\;\label{line:suparm_update}
}
Update
$\usersuparmvec_\usersymbol=\suparmdatamat_\usersymbol^{-1} \suparmlabelvec_\usersymbol, \userarmvec_\usersymbol=\armdatamat_\usersymbol^{-1}\left(\armlabelvec_\usersymbol+(1-\lambda) \usersuparmvec_\usersymbol\right)$\;
Select an {\arm} $\armsymbol_t \in \armset_t$ according to Equation \ref{eq:arm_selection}\;\label{line:arm_selection}
Receive the reward $\reward_{\armsymbol_t, t}$ according to Equation \ref{eq:reward}\;
Update $\armdatamat_\usersymbol=\armdatamat_\usersymbol+\lambda \armvec_{\armsymbol_t} \armvec_{\armsymbol_t}^\top$, $\armlabelvec_\usersymbol=\armlabelvec_\usersymbol+\lambda \reward_{\armsymbol_{t}} \armvec_{\armsymbol_t}$\;\label{line:arm_update}
\caption{{\myalg}}
\label{alg:overview}
\end{algorithm}

Algorithm \ref{alg:overview} describes the proposed {\myalg} which implements Line \ref{line:interactbegin}-\ref{line:framework-forend} of Algorithm \ref{alg:framework}.
Following the work \cite{zhang2020conversational}, we model the {\arm}s and {\suparm}s separately.
Specifically, for user $\usersymbol$ the {\agent} has two sets of estimated features $\userarmvec_\usersymbol$ and $\usersuparmvec_\usersymbol$.
$\userarmvec_\usersymbol$ models the user's preference on {\arm}s, whereas $\usersuparmvec_\usersymbol$ models the user's preference on {\suparm}s.

At each iteration $i$, user $\usersymbol$ with unknown feature $\uservec^*$ will interact with the {\agent}.
The {\agent} then decides whether to conduct conversation based on the user's historical interactions and the system's preset conversational frequency function.
When conducting conversations, {\myalg} selects the pair of {\suparm}s which maximizes error reduction, and queries the user's comparative preference over them.
The received feedback will then be utilized to update $\usersuparmvec_\usersymbol$, which helps the {\agent} to quickly capture users' preference and, as the result, to recommend the user's desired item.
To efficiently incorporate feedback, three update mechanisms are proposed.
We will discuss them thoroughly in Section \ref{sec:alg_conversation}.
Finally, {\myalg} recommends an item to the user, and receives a reward.
The recommendation process relies on a upper-confidence-based strategy.
The detailed methods for item recommendation can be found in Section \ref{sec:alg_recommendation}.

\subsection{Comparison-Based Conversation}\label{sec:alg_conversation}
In this section we will discuss how to conduct comparison-based conversations (Line \ref{line:suparm_selection}-\ref{line:suparm_update} of Algorithm \ref{alg:overview}).
In each conversation, the {\agent} needs to ask a relative question.
Once the user expresses comparative preference over the question, our estimation of the user's parameters can then be updated.
To be specific, two key problems will be discussed in details and conquered: 1) how to select {\suparm}s to conduct comparison-based {\conversation}s (Line \ref{line:suparm_selection} of Algorithm \ref{alg:overview}); and 2) how to update the model effectively with relative feedback (Line \ref{line:suparm_update} of Algorithm \ref{alg:overview}). 

In this work, we propose three variants of the {\myalg} algorithm: \textbf{\updatepos}, \textbf{\updateposneg}, and \textbf{\updatediff}.

\subsubsection{\textbf{\updatepos} and \textbf{\updateposneg}}\label{sec:pos-posneg}
On the first attempt, we adopt an absolute model for relative questions inspired by \cite{christakopoulou2016towards}.
The insight behind the model is that relative feedback inferring the user's comparative preference can be interpreted as two individual observations of {\suparm}.
For example, if the user shows more inclination to comedy movies than sci-fi ones, the {\agent} will translate it into absolute information: 1) the user likes comedy movies; and 2) the user dislikes sci-fi movies.
In this manner, we estimate the {\suparm} level preference of user $\usersymbol$ by:

\begin{equation}\label{eq:suparm_level}
    \usersuparmvec_\usersymbol=\arg \min _{\usersuparmvec} \sum_{(\suparmvec, \feedback) \in \observeset_\usersymbol}\left(\usersuparmvec^\top \suparmvec -\feedback\right)^{2}+\tilde{\lambda}\|\usersuparmvec\|_{2}^{2},
\end{equation}
where $(\suparmvec, \feedback)$ stands for an absolute observation of {\suparm} with $\suparmvec \in \suparmset$ and $\feedback \in \{0, 1\}$, and $\observeset_\usersymbol$ denotes the collected observation dataset of user $\usersymbol$.
Equation \ref{eq:suparm_level} has a closed-form solution $\usersuparmvec_\usersymbol=\suparmdatamat_\usersymbol^{-1} \suparmlabelvec_\usersymbol$, where
\begin{equation*}
\begin{aligned}
    \suparmdatamat_\usersymbol &= \sum_{(\suparmvec, \feedback) \in \observeset_\usersymbol} \suparmvec \suparmvec^\top +\tilde{\lambda} \mathbf{I},\\
    \suparmlabelvec_\usersymbol &= \sum_{(\suparmvec, \feedback) \in \observeset_\usersymbol} \suparmvec \feedback.
\end{aligned}
\end{equation*}

Based on the above model, we formulate two variants of the proposed algorithm \textbf{\updatepos} and \textbf{\updateposneg}.
They use the same mechanism to select relative questions.
Suppose at $t$-th round of interaction with user $\usersymbol$, the {\agent} has access a subset of {\arm}s $\armset_t$ and a subset of {\suparm} $\suparmset_t$. 
The contextual vectors for {\arm}s are denoted by $\armmat_t$, whose rows are composed of $\{\armvec_{\armsymbol}^\top: \armsymbol \in \armset_t\}$.
To collect the most informative observations, we expect to select {\suparm}s that minimize the estimation error $\mathbb{E}\left[\left\|\armmat_{t} \userarmvec_\usersymbol-\armmat_{t} \userarmvec_\usersymbol^*\right\|_{2}^{2}\right]$.
To select the optimal {\suparm} to observe, we can approximately choose:

\begin{equation}\label{eq:best_key-term}
    \suparmsymbol=\arg \max _{k^{\prime} \in \suparmset_t} \frac{\left\|\armmat_t \armdatamat_\usersymbol^{-1} \suparmdatamat_\usersymbol^{-1} \suparmvec_{\suparmsymbol^{\prime}}\right\|_{2}^{2}}{1+\suparmvec_{\suparmsymbol^{\prime}}^{\top} \suparmdatamat_\usersymbol^{-1} \suparmvec_{\suparmsymbol^{\prime}}},
\end{equation}
according to Theorem 2 in \cite{zhang2020conversational}.
The essence behind Equation \ref{eq:best_key-term} is to greedily choose an {\suparm} that could reduce the most uncertainty in the upcoming round of recommendation.

In this manner, we apply the following mechanism to choose the pair of {\suparm}s: 1) we select a {\suparm} $\suparmsymbol_1$ with feature $\suparmvec_{\suparmsymbol_1}$ according to Equation \ref{eq:best_key-term}; 2) then, we conduct a pseudo update by updating the model with $\suparmvec_{\suparmsymbol_1}$ according to Algorithm \ref{alg:suparmupdate}; 3) finally, we select another {\suparm} $\suparmsymbol_2$ according to Equation \ref{eq:best_key-term} with the updated model.


The only difference between two variants is how to interpret relative feedback to update the model or, equivalently, how to construct observation samples.

\begin{algorithm}[t]
\SetAlgoLined
\SetKwInOut{Input}{Input}
\SetKwInOut{Init}{Init}

\Input{user $\usersymbol$ with parameters $\suparmdatamat_\usersymbol$,  $\suparmlabelvec_\usersymbol$; observation $(\suparmvec, \feedback)$}
\BlankLine

$\suparmdatamat_\usersymbol = \suparmdatamat_\usersymbol + \suparmvec \suparmvec^\top$\;
$\suparmlabelvec_\usersymbol = \suparmlabelvec_\usersymbol + \feedback \suparmvec$\;
\caption{{\firstuppersuparm} Level Updating}
\label{alg:suparmupdate}
\end{algorithm}

Suppose during the conversation at round $t$, user $\usersymbol$ expresses that $\suparmsymbol_1$ is preferred rather than $\suparmsymbol_2$.
The {\agent} will therefore receive relative feedback $\feedback_{\suparmsymbol_1, \suparmsymbol_2, t} = 1$.
Then the {\agent} will update the model for user $\usersymbol$ according to Algorithm \ref{alg:suparmupdate}.
For \textbf{\updatepos}, it incorporates only positive information $(\suparmvec_{\suparmsymbol_1}, 1)$.
For \textbf{\updateposneg}, the relative feedback is interpreted as two observations of absolute feedback: a positive observation $(\suparmvec_{\suparmsymbol_1}, 1)$ for the preferred {\suparm} and a negative observation $(\suparmvec_{\suparmsymbol_2}, 0)$ for the less preferred {\suparm}.

\subsubsection{\textbf{\updatediff}}\label{sec:difference}
The above updating methods deal with relative feedback the same as with absolute feedback, which may not fully utilize the advantages of relative feedback.
To improve, we consider training a linear classifier directly on paired data, and develop the updating method \textbf{\updatediff}.

We still apply the estimation of $\usersuparmvec_\usersymbol$ according to Equation \ref{eq:suparm_level}.
But for \textbf{\updatediff}, we incorporate the relative feedback as a single observation $(\suparmvec_{\suparmsymbol_1} - \suparmvec_{\suparmsymbol_2}, 1)$. 
This is essentially training a linear classifier of feature difference using ridge regression.
Note that the unknown parameters can not be fully identified by relative feedback, as $\uservec^*$ and $2 \uservec^*$ induce the same comparison results for any pair $\suparmvec_{\suparmsymbol_1}, \suparmvec_{\suparmsymbol_2}$ in expectation.
Rather, what the linear classifier actually estimates is the normalized feature vector $\frac{\uservec^*}{\|\uservec^*\|_2}$. 
On the other hand, in the {\arm}-level recommendation, the {\agent} learns the norm of the feature $\|\uservec^*\|_2$.
In this manner, {\myalg} learns users' preferences effectively in the sense of both absolute correctness and relative relationship.

To design relative questions, we follow the maximum error reduction strategy which leads us to choose the pair based on the feature difference $\Delta\suparmvec_{\suparmsymbol_1, \suparmsymbol_2} = \suparmvec_{\suparmsymbol_1} - \suparmvec_{\suparmsymbol_2}$:

\begin{equation}\label{eq:best_diff}
    \suparmsymbol_1, \suparmsymbol_2=\arg \max _{\suparmsymbol_1, \suparmsymbol_2 \in \suparmset_t}
    \frac{\left\|\armmat_t \armdatamat_\usersymbol^{-1} \suparmdatamat_\usersymbol^{-1} \Delta\suparmvec_{\suparmsymbol_1, \suparmsymbol_2}\right\|_{2}^{2}}{1+\Delta\suparmvec_{\suparmsymbol_1, \suparmsymbol_2}^{\top} \suparmdatamat_\usersymbol^{-1} \Delta\suparmvec_{\suparmsymbol_1, \suparmsymbol_2}}.
\end{equation}

The {\suparm} selection strategy described above induces the standard \textbf{\updatediff} variant. However, since the number of {\suparm}s can be extremely large, the strategy may not be practical as it has a quadratic time complexity.
To eschew the problem, we further propose an optimized variant called \textbf{\updatediff} (\textit{fast}). When selecting {\suparm}s, it first selects a {\suparm} $\suparmsymbol_1$ by Equation \ref{eq:best_key-term}.
Then, with the first {\suparm} fixed, it chooses another {\suparm} $\suparmsymbol_2$ based on Equation \ref{eq:best_diff}.
In Section \ref{sec:experiments}, we will compare these two variants, showing that \textbf{\updatediff} (\textit{fast}) only suffers little decrease of performance.

\subsection{Item Recommendation}\label{sec:alg_recommendation}
The {\arm}-level recommendation (Line \ref{line:arm_selection}-\ref{line:arm_update} of Algorithm \ref{alg:overview}) is essentially similar to that of LinUCB \cite{li2010contextual}.
The difference is that, to accelerate the process of {\arm} preference learning, the {\agent} utilizes $\usersuparmvec_{\usersymbol}$ as a guidance of $\userarmvec_{\usersymbol}$ by penalizing the difference between $\userarmvec_{\usersymbol}$ and $\usersuparmvec_{\usersymbol}$:

\begin{equation*}
    \userarmvec_{\usersymbol}=\arg \min _{\userarmvec} \lambda \sum_{\tau=1}^{t_\usersymbol}\left(\armvec_{\armsymbol_{\tau}}^{\top} \userarmvec - \reward_{\armsymbol_{\tau}, \tau}\right)^{2}+(1-\lambda)\left\|\userarmvec-\usersuparmvec_{\usersymbol}\right\|_2^{2}.
\end{equation*}

The hyper-parameter $\lambda$ balances {\arm}-level and {\suparm} level learning.
The closed-form solution for $\userarmvec_\usersymbol$ is
$\armdatamat_\usersymbol^{-1}\left(\armlabelvec_\usersymbol+(1-\lambda) \usersuparmvec_\usersymbol\right)$, where 
\begin{equation*}
\begin{aligned}
    \armdatamat_\usersymbol &=\lambda \sum_{\tau=1}^{t_\usersymbol} \armvec_{\armsymbol_{\tau}} \armvec_{\armsymbol_{\tau}}^{T}+(1-\lambda) \mathbf{I}\,,\\
    \armlabelvec_\usersymbol &= \lambda \sum_{\tau=1}^{t_\usersymbol} \armvec_{\armsymbol_{\tau}} \reward_{\armsymbol_{\tau}, \tau}.
\end{aligned}
\end{equation*}

To balance exploration and exploitation, {\myalg} selects an arm to recommend in a upper-confidence based strategy:

\begin{equation}\label{eq:arm_selection}
    \armsymbol_{t}=\arg \max _{\armsymbol \in \armset_{t}} \armvec_{\armsymbol}^\top \uservec_\usersymbol+\lambda \alpha\left\|\armvec_{\armsymbol}\right\|_{\armdatamat_\usersymbol^{-1}} + (1 - \lambda) \tilde{\alpha}\left\| \armdatamat_\usersymbol^{-1}\armvec_{\armsymbol}\right\|_{\suparmdatamat_\usersymbol^{-1}},
\end{equation}
where $\alpha, \tilde{\alpha} \in \R$ are hyper-parameters, and $\|\armvec\|_{\armdatamat}=\sqrt{\armvec^{\top} \armdatamat \armvec}$.

The strategy exploits the current estimation on the reward of {\arm} $\armsymbol$ (the first term of Equation \ref{eq:arm_selection}).
A higher value of the first term indicates that in the previous interaction user $\usersymbol$ has expressed more inclination to {\arm} $\armsymbol$.
On the other hand, in order to eschew being misled by wrong perception, the strategy needs to choose some sub-optimal {\arm}s for exploration.
This is achieved by including uncertainty estimation (the second and third term of Equation \ref{eq:arm_selection}).
The second term estimates the uncertainty in the {\arm}-level, while the third term estimates the uncertainty from {\suparm} level parameters.
Please refer to \cite{zhang2020conversational} for more insights.

\subsection{Feedback Sharing}
In the former sections, we have made no assumption on users' features.
However, in practice, groups of users usually share some similar preferences.
A rich body of methods has been proposed to model and exploit users' similar behaviors such as collaborative filtering \cite{das2007google}.
We take a step further and claim that, compared to absolute preferences, comparative preferences could be more consistent among users.
The reason is that user biases can be eliminated from the feedback when making comparisons.
Therefore, no matter how biased each user is, we can still consider users with similar comparative preferences are of the same type.

Based on the hypothesis, we propose to utilize the similarity by feedback sharing.
Suppose the {\agent} interacts with user $\usersymbol$ at iteration $i$.
Once we receive relative feedback from $\usersymbol$, we can update parameters of the similar users all at once:

\begin{equation*}
    \begin{aligned}
        \suparmdatamat_\usersymbol 
        &= \suparmdatamat_\usersymbol + \suparmvec \suparmvec^\top,\\
        \suparmlabelvec_\usersymbol &= \suparmlabelvec_\usersymbol + \feedback \suparmvec, \forall \usersymbol \in \userset.
    \end{aligned}
\end{equation*}
In this manner, conversation and recommendation actually form a two-stage learning process.
When conducting conversations, the {\agent} learns common preferences of a group of users.
In the meantime, the received rewards from item recommendation imply users' individual bias.
In section \ref{sec:experiments}, we will compare the performance of sharing absolute feedback and relative feedback.
We leave more comprehensive study of sharing mechanism as future work.

\section{Experiments}\label{sec:experiments}
To validate our proposed {\myalg} algorithm empirically, we evaluate and compare the performances of several state-of-the-art algorithms on both synthetic and real-world datasets\footnote{The source code is available at \url{https://github.com/fffffarmer/RelativeConUCB}.}.

\subsection{Experimental Setup}
In this section we describe our experimental setup.
We start with the baselines to compare.
Then we describe how the simulation environment is constructed and the evaluation metrics.
Finally, we list several research questions that we seek to answer.

\subsubsection{Baselines}
The following algorithms are selected as the representative baselines to be compared with:

\begin{itemize}
    \item \textbf{LinUCB} \cite{li2010contextual}: A state-of-the-art non-conversational contextual bandit approach.
    \item \textbf{ConUCB} \cite{zhang2020conversational}: A recently proposed conversational contextual bandit approach.
    When conducting conversations with user $\usersymbol$, it chooses the {\suparm} $\suparmsymbol$ which maximizes error reduction, queries the user's preference on it, and receives absolute feedback $\feedback \sim \bernoulli(\suparmvec_{\suparmsymbol}^{\top} \uservec_\usersymbol^{*})$.
\end{itemize}

For LinUCB, we set hyper-parameters $\lambda = 1, \alpha=0.5$. For both ConUCB and {\myalg}, we set hyper-parameters $\lambda=0.5$, $\tilde{\lambda} = 1$, $\sigma=0.05$, $\alpha = 0.25$, $\tilde{\alpha} = 0.25$ according to the default configuration in ConUCB's public source code.

\subsubsection{Simulation} For all experiments, we sequentially simulate $N=400 \times |\userset|$ iterations.
At each iteration $i$, the incoming {\user} $\usersymbol$ with $t = t_\usersymbol$ rounds of historical interaction is chosen randomly from the user set $\userset$.
For item recommendation, the {\agent} can select one out of 50 {\arm}s in the candidate set $\armset_t$, which is randomly drawn from $\armset$ without replacement.
The {\suparm} candidate set $\suparmset_t$ contains those {\suparm}s that are related to any {\arm} in $\armset_t$.
Furthermore, we set the conversation frequency function $b(t) = 5 \lfloor\log (t)\rfloor$ which induces more conversations at the early stage to quickly capture preferences \cite{zhang2020conversational}.
The reward for {\arm}s and relative feedback for {\suparm}s are generated by Equation \ref{eq:reward} and Equation \ref{eq:relative_feedback} where $\sigma_g = 0.1$.
The distribution $p$ over users is set uniform, suggesting that each user will averagely interact with the {\agent} for 400 rounds.
The final results are reported by conducting 10 repeated runs.

\subsubsection{Metrics}
We mainly use \textit{cumulative regret} in Equation \ref{eq:overall_cum_regret} to measure the performance of algorithms.
As a widely used metric in the multi-armed bandit scenario, it captures the degree of satisfaction the user feels towards recommendation in a long run.
Besides, in Section \ref{sec:synthetic_exp}, we report \textit{averaged reward} over iterations $A(N) \defeq \frac{1}{N} \sum_{i=1}^N \reward_i$
to help clearly show the gap between different algorithms.

\subsubsection{Research Questions}
The designed experiments aim to answer the following evaluation questions:

\begin{itemize}
    \item[\textbf{RQ1.}] Can relative feedback be incorporated well to help preference learning, and which variant of our algorithm is more effective?
    \item[\textbf{RQ2.}] Can we observe the benefit brought from feedback sharing?
    \item[\textbf{RQ3.}] Under which circumstances will relative feedback be more superior to absolute feedback?
\end{itemize}

\begin{table}[b!]
    \centering
    \caption{Statistics of the datasets used in our experiments.}
    \label{tab:dataset_stats}
    \begin{tabular}{llll}
        \toprule
         & Synthetic & Last.FM & Movielens \\
         \midrule
         \# of {\suparm}s & 500 & 2,726 & 5,585 \\
         avg. \# of related {\suparm}s & 0.521 & 16.55 & 11.37 \\
         avg. \# of related items & 5.214 & 12.14 & 4.071 \\
         \bottomrule
    \end{tabular}
\end{table}

\begin{figure}[t!]
    \centering
    \includegraphics[width=0.99\linewidth]{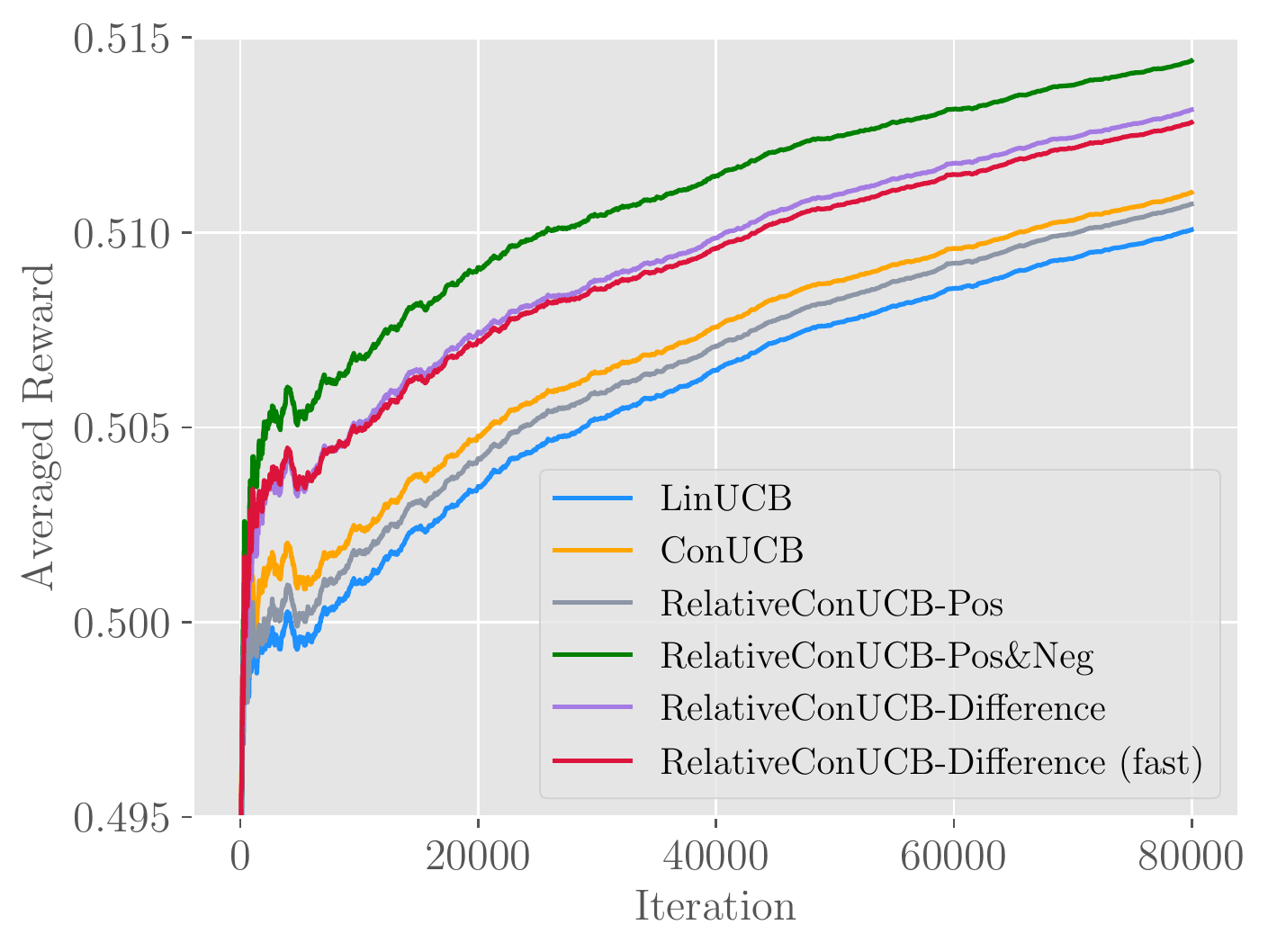}
    \caption{Averaged reward on synthetic data.}
    \Description{A figure demonstrating the proposed algorithm can outperform the baselines on synthetic data.}
    \label{fig:synthetic}
\end{figure}

\subsection{Synthetic Data}\label{sec:synthetic_exp}
To answer \textbf{RQ1}, we first evaluate the algorithms on the synthetic data. 
\subsubsection{Data Generation} We consider a setting of a user set $\userset$ with $|\userset| = 200$, an {\arm} set $\armset$ with $|\armset| = 5,000$ and a {\suparm} set $\suparmset$ with $|\suparmset| = 500$.
The relation matrix $\weightmat \in \R^{|\armset| \times |\suparmset|}$ between {\arm}s and {\suparm}s is constructed by the following procedures: 
1) for each {\suparm} $\suparmsymbol$, we uniformly sample a subset of $n_\suparmsymbol$ related {\arm}s $\armset_\suparmsymbol \subseteq \armset$ without replacement, where $n_\suparmsymbol$ is an integer drawn uniformly from $[1, 10]$; 
2) we assign $\weightmat_{\armsymbol, \suparmsymbol} = 1 / n_\armsymbol$, assuming {\arm} $\armsymbol$ is related to a subset of  $n_\armsymbol$ {\suparm}s $\suparmset_\armsymbol$.

To generate the arms' feature vector, we follow similar procedures in \cite{zhang2020conversational}:
1) we first sample a pseudo feature in $d - 1$ dimension $\suparmpseudovec_\suparmsymbol \sim \gaussian\left(\mathbf{0}, \sigma^{2} \mathbf{I}\right)$ for each $\suparmsymbol \in \suparmset$, where $d = 50$ and $\sigma=1$;
2) for each arm $\armsymbol$, its feature vector $\armvec_\armsymbol$ is drawn from $\gaussian\left(\sum_{k \in \suparmset_{\armsymbol}} \suparmpseudovec_{\suparmsymbol} / n_{\armsymbol}, \sigma^{2}\mathbf{I}\right)$;
3) finally, feature vectors are normalized and added one more dimension with constant $1$: $\armvec_\armsymbol \gets (\frac{\armvec_\armsymbol}{\|\armvec_\armsymbol\|_2}, 1)$.

For the ground-truth feature of each user $\usersymbol$, we first randomly sample a feature vector $\uservec_\usersymbol^* \in \R^{d - 1}$ from $\gaussian\left(\mathbf{0}, \sigma^{2} \mathbf{I}\right)$.
Then we introduce two parameters: $\scale_\usersymbol$ and $\bias_\usersymbol$
They are drawn uniformly from $[0, 0.5]$ and $[\scale_\usersymbol, 1 - \scale_\usersymbol]$ respectively, controlling the variance and mean of expected reward.
The final feature vector of user $\usersymbol$ is $\uservec_\usersymbol^* \gets (\scale_\usersymbol \frac{\uservec_\usersymbol^*}{\|\uservec_\usersymbol^*\|_2}, \bias_\usersymbol)$.
Note that the normalization step on both item features and user features aims to bound the expected reward $\armvec_\armsymbol^\top\uservec_\usersymbol^* \in [0, 1]$, $\forall \armsymbol \in \armset, \usersymbol \in \userset$.

\subsubsection{Results} The results are given in Figure \ref{fig:synthetic}, which plots the averaged reward over 80,000 iterations.
We can first notice that all other algorithms outperform LinUCB, indicating the advantage of conducting conversations.
The improvements are mainly reflected in the early stage, when the {\agent} frequently converses with users.

\paragraph{Answer to \textbf{RQ1.}}
We can clearly observe that variant \textbf{\updateposneg} and \textbf{\updatediff} have significant improvements over ConUCB.
\textbf{\updatepos}, on the other hand, performs similarly to LinUCB.

Besides, the performances of \textbf{\updatediff} (\textit{fast}) and \textbf{\updatediff} are very close.
This validates that \textbf{\updatediff} (\textit{fast}), as an approximation of \textbf{\updatediff}, can significantly reduce time complexity with little performance decrease.

\begin{figure*}[h!]
    \centering
    \begin{minipage}[b]{0.245\textwidth}
      \subfigure[Last.FM (without feedback sharing)]{\includegraphics[width=\textwidth]{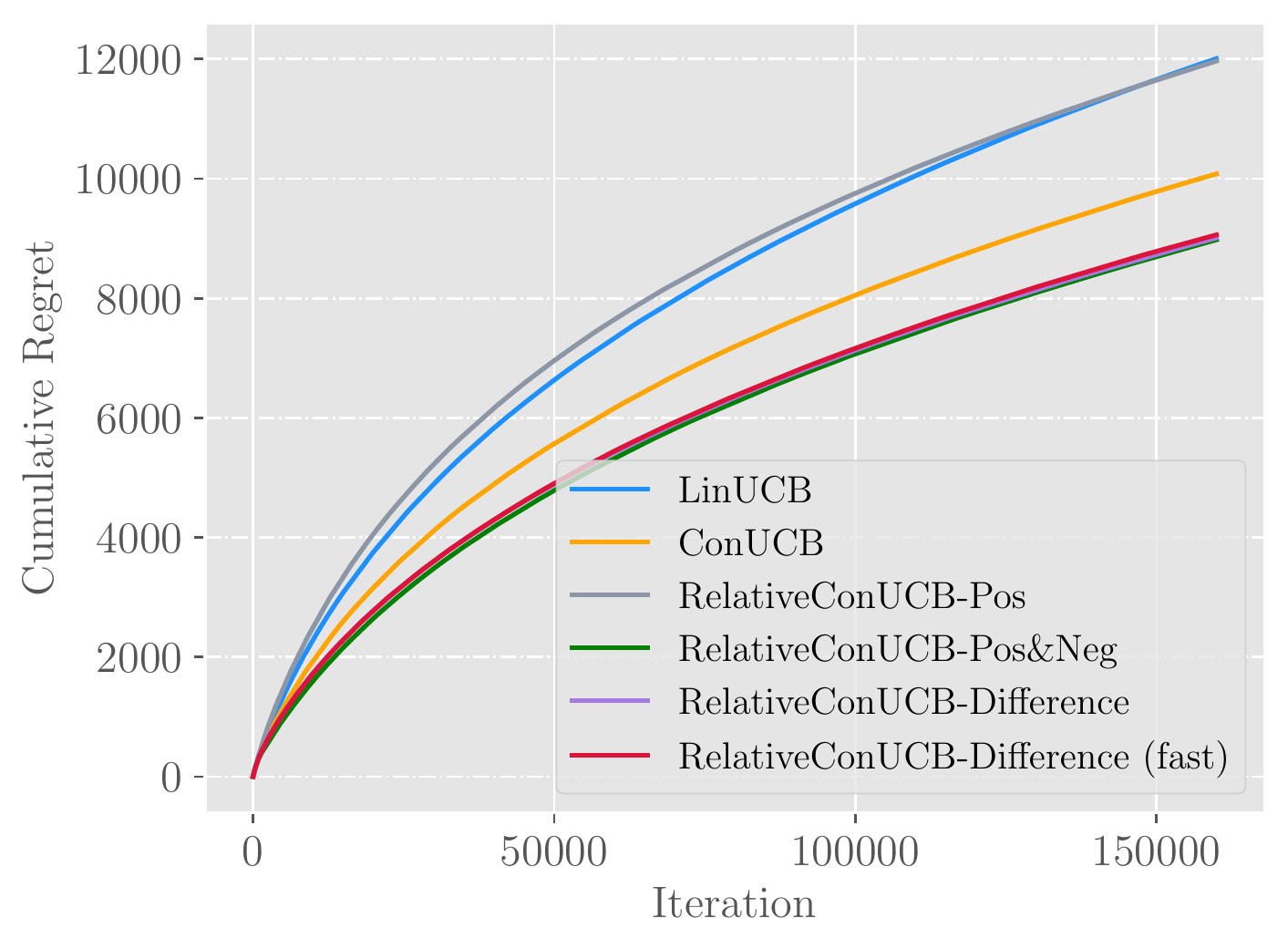}\label{fig:real_lastfm_without}}
    \end{minipage}
    \begin{minipage}[b]{0.245\textwidth}
      \subfigure[Movielens (without feedback sharing)]{\includegraphics[width=\textwidth]{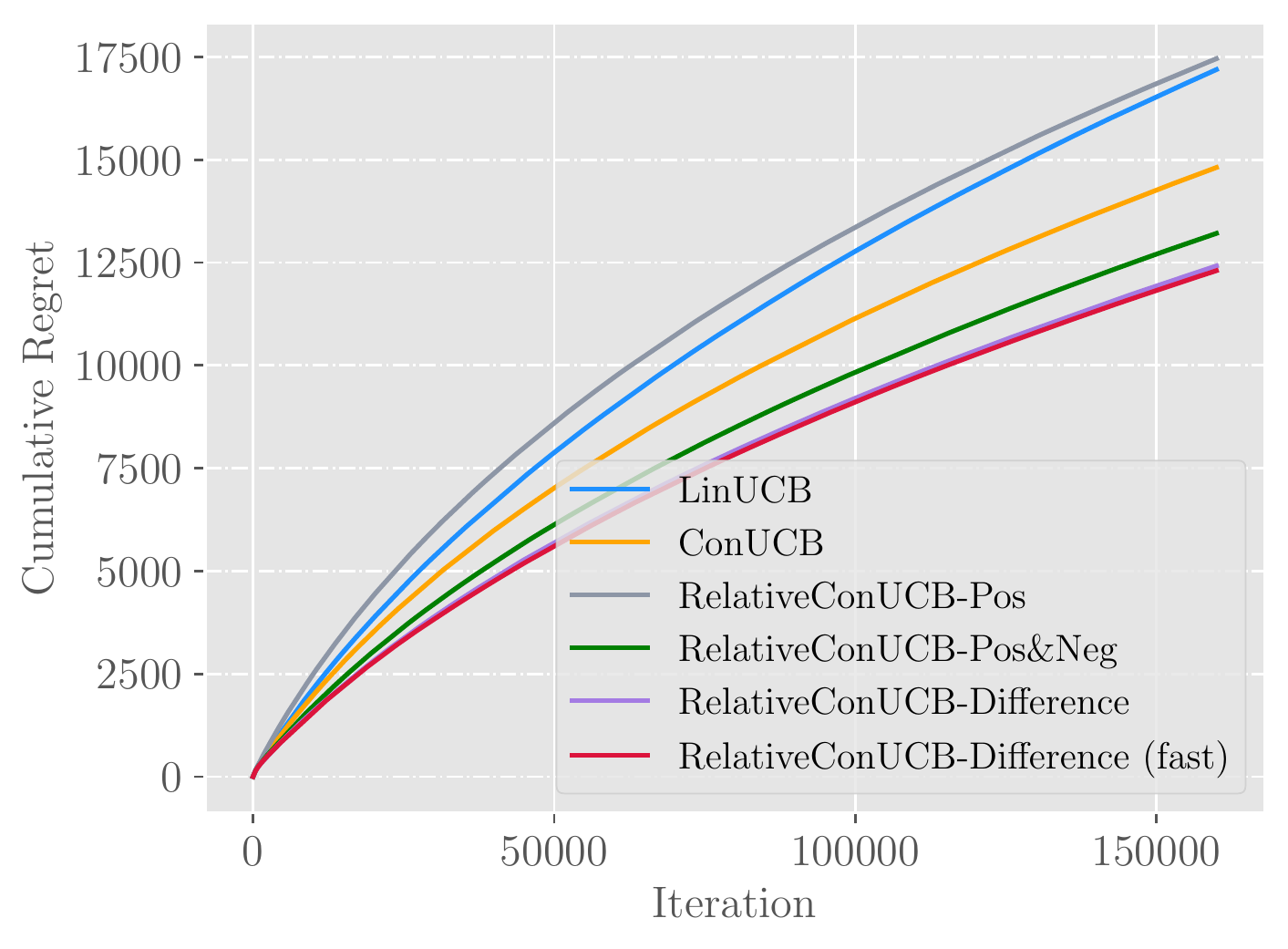}\label{fig:real_movielens_without}}
    \end{minipage}
    \begin{minipage}[b]{0.245\textwidth}
      \subfigure[Last.FM (with feedback sharing)]{\includegraphics[width=\textwidth]{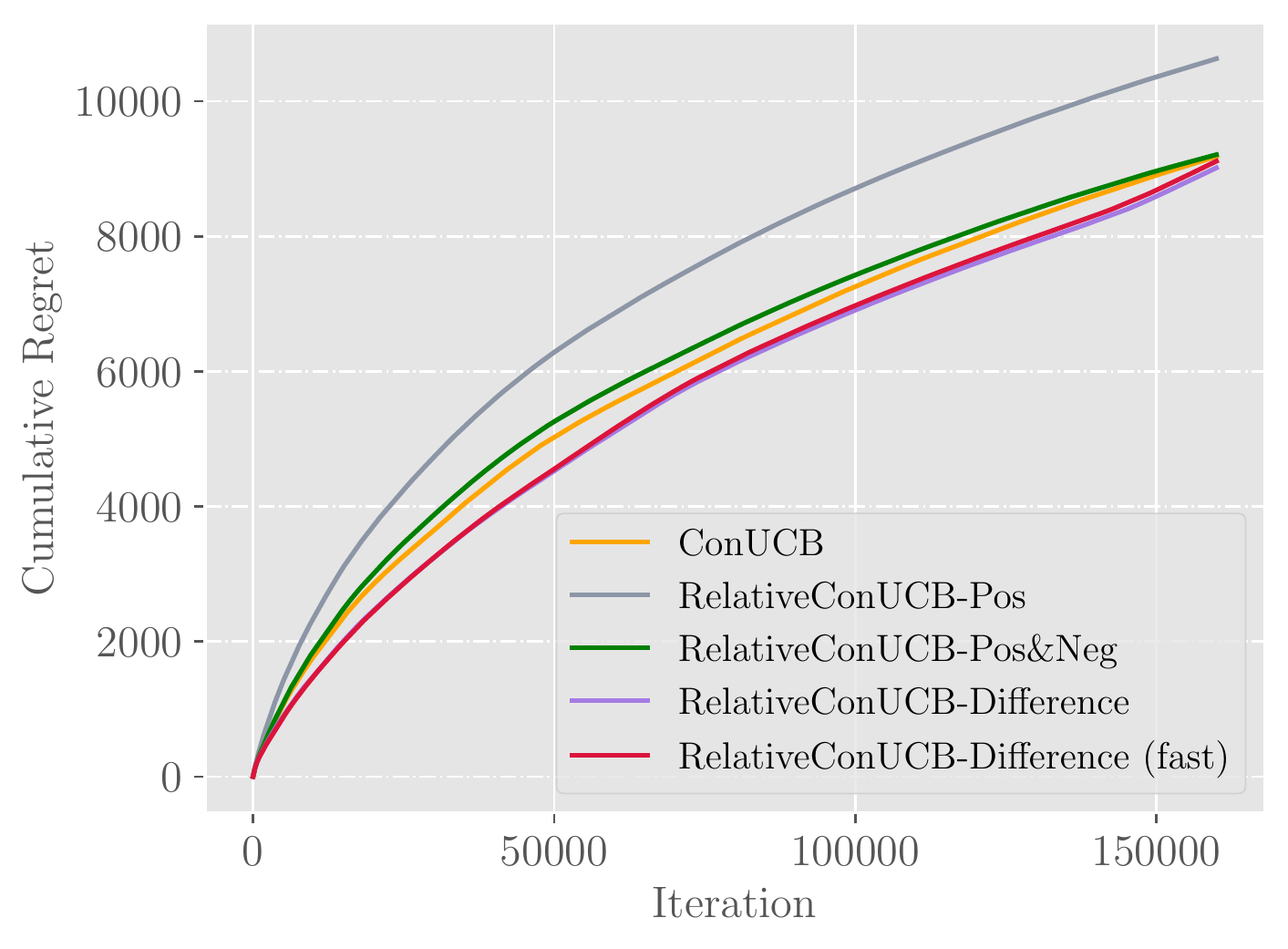}\label{fig:real_lastfm_with}}
    \end{minipage}
    \begin{minipage}[b]{0.245\textwidth}
      \subfigure[Movielens (with feedback sharing)]{\includegraphics[width=\textwidth]{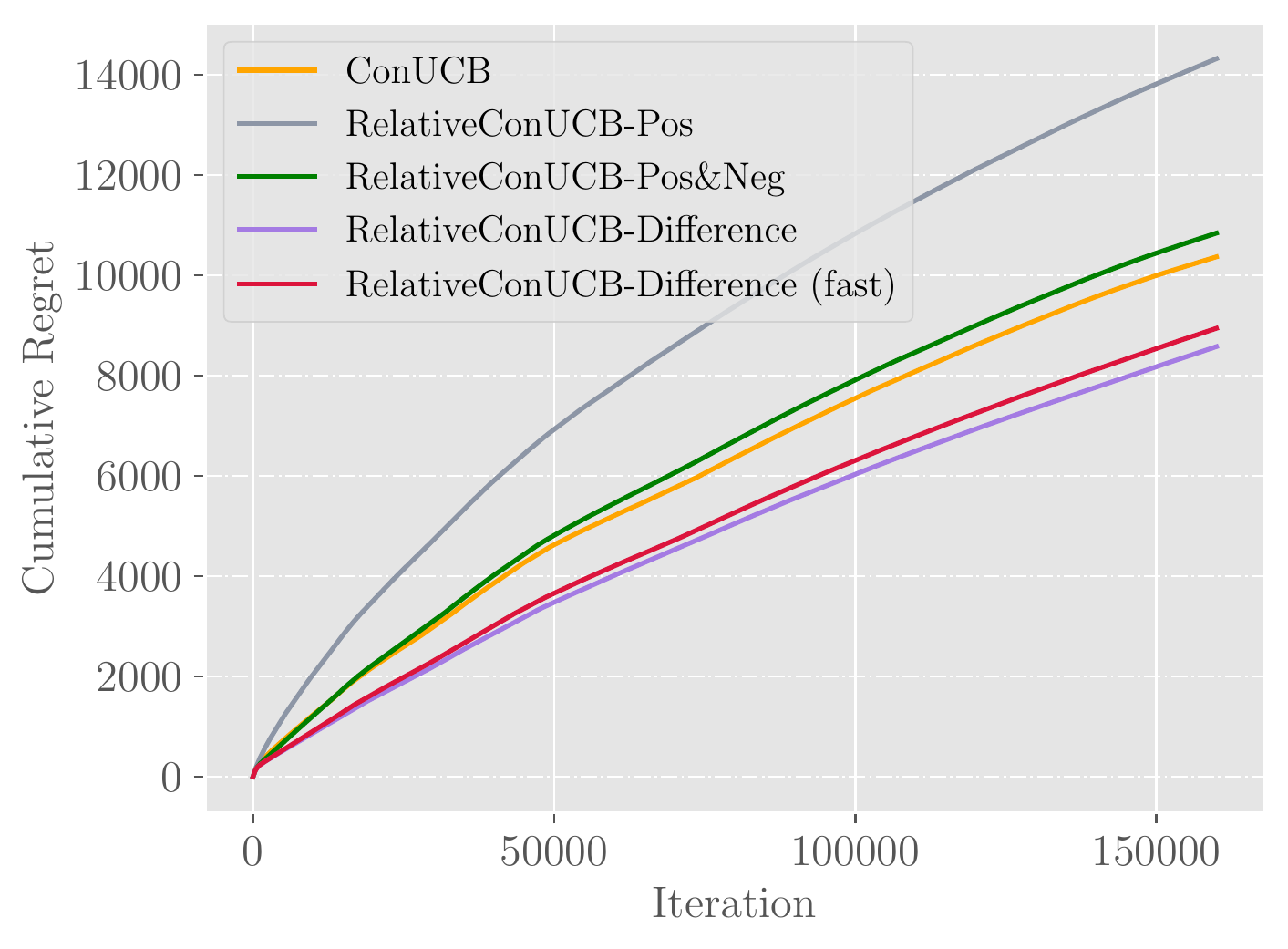}\label{fig:real_movielens_with}}
    \end{minipage}
    \caption{Cumulative regret on real-world datasets.}
    \label{fig:real} 
    \Description{Figures of experimental results on real datasets. They verify the conclusion of synthetic experiment, and show that feedback sharing can not directly bring significant improvements.}
\end{figure*}

\subsection{Real-world Datasets}
We next evaluate {\myalg} on two real-world datasets from different domains, which are released in \cite{cantador2011second}: Last.FM\footnote{\url{http://www.lastfm.com}} for music artist recommendation and Movielens\footnote{\url{http://www.grouplens.org}} for movie recommendation.
Last.FM dataset contains 186,479 interaction records between 1,892 users and 17,632 artists.
Movielens dataset, which extends the original MovieLens10M dataset to include tagging information in IMDb\footnote{\url{http://www.imdb.com}} and Rotten Tomatoes\footnote{\url{http://www.rottentomatoes.com}}, contains 47,957 interaction records between 2,113 users and 10,197 movies.
Artists and movies are treated as items in our setting.
We then infer users' real feedback on items based on the interaction records: if the user has assigned attributes to the item, the feedback is 1, otherwise the feedback is missing.
For each dataset, we extract 2,000 items with most assigned attributes and 500 users who have assigned most attributes.
For each item, we only leave at most 20 attributes that are associated with most items, and consider them as the related key-terms of the item.
The statistics of processed data is shown in Table \ref{tab:dataset_stats}, where the average number of related {\suparm}s for items and the average number of related items for {\suparm}s are also reported.

\subsubsection{Data Generation}\label{sec:real_generation}
Before we start, we randomly choose 100 users and formulate a binary matrix $H \in \{0, 1\}^{100 \times 2\text{k}}$ which stands for the interaction history.
We derive feature vectors in $d - 1$ ($d = 50$) dimension for items by applying one-class matrix factorization \cite{chin2016libmf} on $H$, and then normalize them by $\armvec_\armsymbol \gets (\frac{\armvec_\armsymbol}{\|\armvec_\armsymbol\|_2}, 1)$, $\forall \armsymbol \in \armset$.
The constructed features will be used to conduct experiments on the records of the remaining 400 users.

For missing feedback of the remaining users, we again apply the matrix factorization technique to reconstruct the feedback matrix $F \in \{0, 1\}^{400 \times 2\text{k}}$.
Next, in order to derive the ground truth user features, we use a simple autoencoder \cite{rumelhart1985learning} with four hidden layers and ReLU activation function.
The autoencoder takes the user's feedback (i.e., the corresponding row of $F$) as its input, encodes it into a latent representation $\uservec_\usersymbol^* \in \R^{d - 1}$, and normalizes it by $\uservec_\usersymbol^* \gets (\frac{\uservec_\usersymbol^*}{\|\uservec_\usersymbol^*\|_2}, 1) / 2$.
Finally, we reconstruct the feedback by $\armvec_\armsymbol^\top \uservec_\usersymbol^*$.

When training, we use MSE loss as the criterion and Adam with learning rate $3 \times 10^{-4}$ as the optimizer.
The training runs 200 epochs.

\subsubsection{Results} In this experiment, we compare the performance of algorithms both in the original setting and the feedback sharing setting.
For ConUCB, it shares conversational absolute feedback to all users.
For {\myalg}, relative feedback is shared instead.
The experimental results on real-world dataset are shown in Figure \ref{fig:real}.

\paragraph{Answer to \textbf{RQ1.}}
We first evaluate the results with feedback sharing disabled.
In Figure \ref{fig:real_lastfm_without} and \ref{fig:real_movielens_without}, we again observe the advantage of using relative feedback, comparing all algorithms in terms of cumulative regret.
On Last.FM dataset, \textbf{\updatediff} (\textit{fast}) and \textbf{\updateposneg} outperforms ConUCB by $10.12\%$ and $10.93\%$ respectively.
On Movielens dataset, \textbf{\updatediff} (\textit{fast}) and \textbf{\updateposneg} outperforms ConUCB by $16.93\%$ and $10.82\%$ respectively.

\paragraph{Answer to \textbf{RQ2.}}
As illustrated in 
Figure \ref{fig:real_lastfm_with} and \ref{fig:real_movielens_with}, directly sharing relative feedback among all users may not result in good performance.
Compared to ConUCB, \textbf{\updatediff} and \textbf{\updateposneg} do not show any advantage, which suggests that feedback sharing needs to be conducted more carefully by only considering similar users.

\subsection{Drifted Real-world Datasets}\label{sec:drift_exp}
As previously shown, sharing feedback among various kinds of users can not generally improve the performance of {\myalg}.
But what if we only share feedback among similar users?
Can user biases be effectively eliminated by using relative feedback instead of absolute one?

To take a step further, we design a drifted environment based on the real-world datasets.
It models the users to be in some sense similar, sharing common comparative preferences.
But individuals have their own bias, viewing items with different criteria.
Specifically, for any two items, they may agree on which one is preferred.
But the absolute attitude (like or dislike) towards each of them can be quite different among the users, depending on how critical each user is.
\subsubsection{Data Generation}
In the drifted environment, we assume the users' comparative preferences are inherently the same.
To be specific, we model the feature vectors of user to be different only in the bias term.
Following similar procedures in Section \ref{sec:real_generation}, we apply a four-layer autoencoder with two trainable embeddings $\uservec_{base} \in \R^{d-1}$ and $c \in [0, 0.5]$ to construct drifted user features.
$\uservec_{base}$ aims to extract the common preferences of users and $c$ controls the variance of expected reward.
Taking the user's feedback as the input, the autoencoder outputs a real number $\tau \in [0, 1]$.
The user feature is then constructed by $\uservec_\usersymbol^* \gets (\scale \frac{\uservec_{base}}{\|\uservec_{base}\|_2}, \bias_\usersymbol)$, where
$\bias_\usersymbol = \scale + (1 - \scale) \tau$.
In this manner, the autoencoder constructs a drifted user group where preferences only differ in $\bias_u$.

When training on the autoencoder, we use MSE loss as the criterion and Adam with learning rate $3 \times 10^{-3}$ as the optimizer.
The training runs 1,000 epochs.

\subsubsection{Results} 
\begin{figure}[h!]
    \centering
    \begin{minipage}[b]{0.49\linewidth}
      \subfigure[Last.FM (with feedback sharing)]{\includegraphics[width=\textwidth]{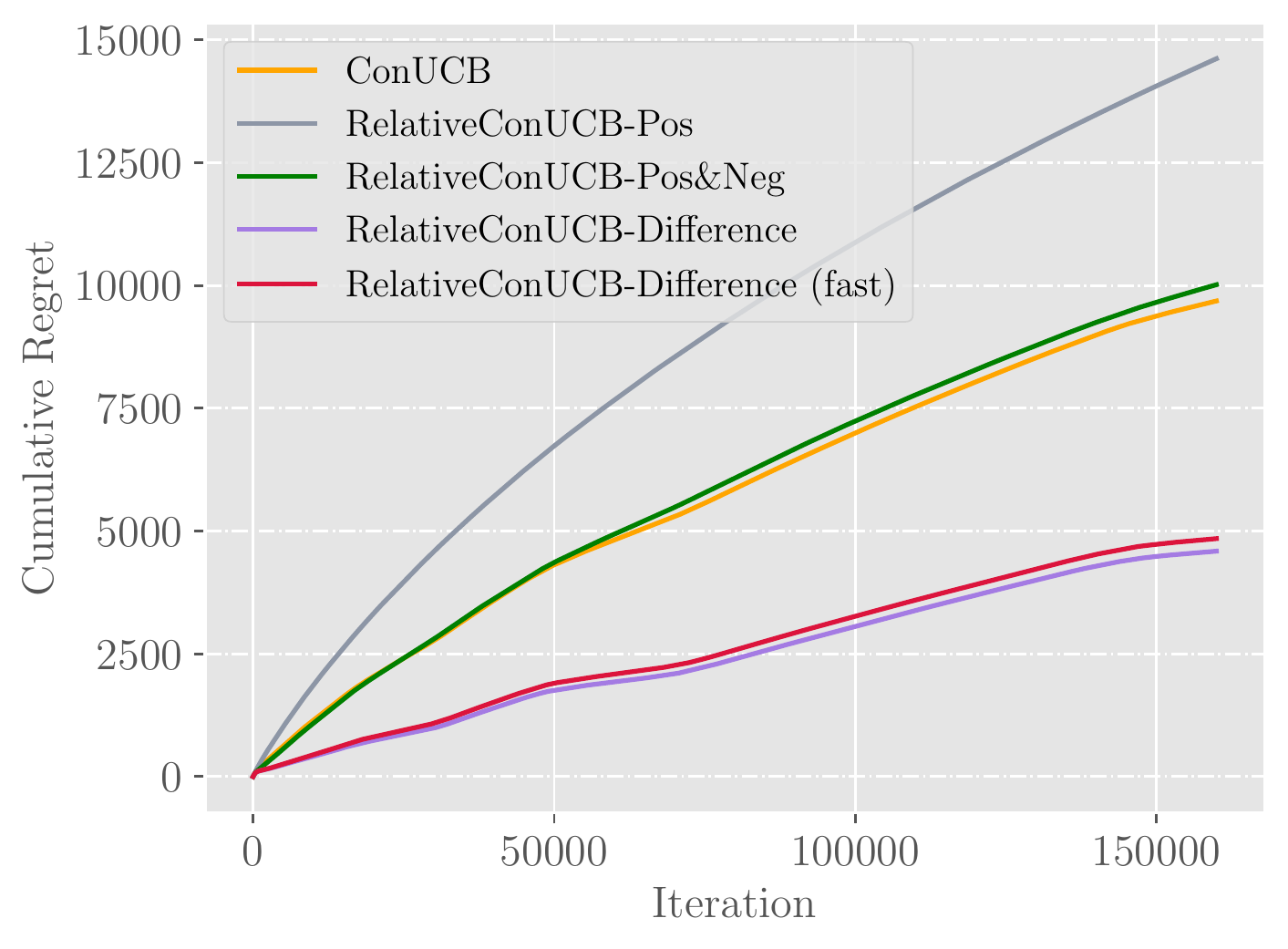}}
    \end{minipage}
    \begin{minipage}[b]{0.49\linewidth}
      \subfigure[Movielens (with feedback sharing)]{\includegraphics[width=\textwidth]{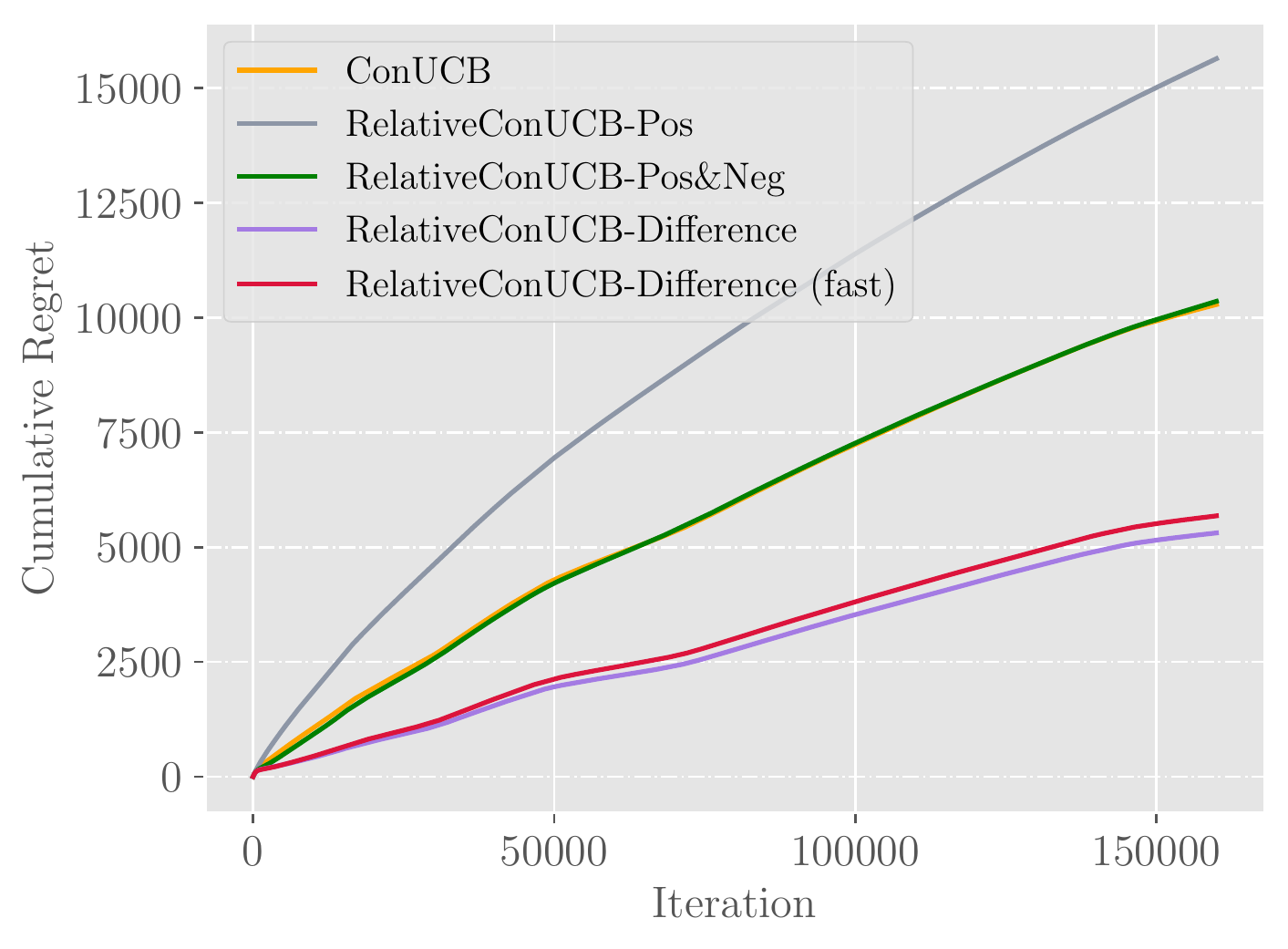}}
    \end{minipage}
    \caption{Cumulative regret on real-world datasets in a drifted setting.}
    \label{fig:real_drift} 
    \Description{Figures presenting results on drifted datasets. In an idealized environment, we can observe significant improvements brought from feedback sharing.} 
\end{figure}

The results are shown in Figure \ref{fig:real_drift}.
\paragraph{Answer to \textbf{RQ1.}}
As illustrated in the figures, \textbf{\updatediff} outperforms other algorithms by a substantial degree.
Based on the result, we argue that \textbf{\updatediff} can better capture the intrinsic common preferences of users.
\paragraph{Answer to \textbf{RQ2.}}
As we can observe, among a group of similar users, sharing relative feedback can bring significant improvements.
Compared to ConUCB, \textbf{\updatediff} (\textit{fast}) outperforms by $49.96\%$ and $44.75\%$ on Last.FM and Movielens dataset respectively. 
The result demonstrates that relative feedback can conquer the problem of user bias, which makes it more effective to utilize user feedback.

\subsection{Different User Groups}\label{sec:synthetic}
In Section \ref{sec:drift_exp}, we have observed that when users have similar but biased preferences, relative feedback can be greatly utilized by sharing among users to better help recommendation.
To explore the superiority in a fine-grained prospective, we setup another set of experiments to answer \textbf{RQ3}.
\subsubsection{Data Generation} 
We modify the procedures of user feature generation based on the synthetic experiment.
For each user $\usersymbol$, we consider the ground-truth feature is composed of three parts: users' common feature $\uservec_{base}$, the user's individual feature $\uservec_\usersymbol$\footnote{It is a slight abuse of notion, as $\uservec_\usersymbol$ usually stands for the estimated user feature.}, and the user's bias $b_\usersymbol$.
$\uservec_{base}$ and $\uservec_\usersymbol$ are both in $d - 1$ dimension, and are drawn randomly from $\gaussian\left(\mathbf{0}, \sigma^{2} \mathbf{I}\right)$.
We introduce two parameters $\scale \in [0, 0.5]$ and $\individual \in [0, +\infty)$ to control the reward variance and user similarity.
The final feature vector of user $\usersymbol$ is $\uservec_\usersymbol^* \gets (\scale \frac{\uservec_{base} + \individual \uservec_\usersymbol}{\|\uservec_{base} + \individual \uservec_\usersymbol\|_2}, \bias_\usersymbol)$, where $\bias_\usersymbol$ is drawn uniformly from $[\scale, 1 - \scale]$.
In this experiment we set $\scale = 0.4$.

\begin{figure}[t!]
    \centering
    \includegraphics[width=0.99\linewidth]{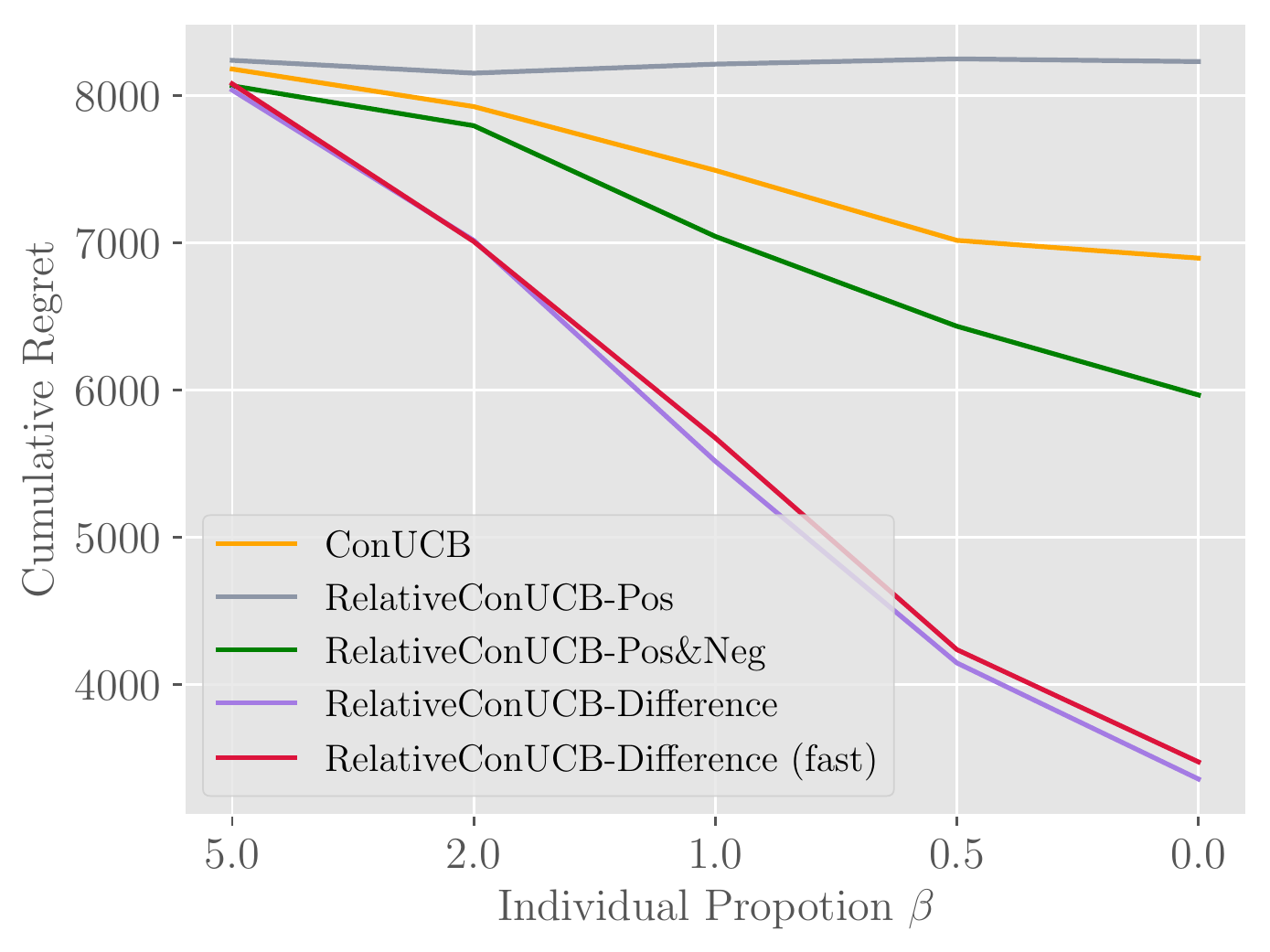}
    \caption{Effect of feedback sharing on different user groups.}
    \label{fig:synthetic_individual_with}
    \Description{A figure demonstrating the trend of performance as users become similar. This helps validate the advantage of relative feedback.}
\end{figure}


\subsubsection{Results} 
Figure \ref{fig:synthetic_individual_with} shows the outcome of the experiment, a comparison of the performances of algorithms under different user individual partitions (i.e., $\individual$).
As the value of $\individual$ gets smaller, users share more similarity in preferences.
Meanwhile, the scale of user bias over the user group remains the same.


\paragraph{Answer to \textbf{RQ3.}} 
From Figure \ref{fig:synthetic_individual_with} we can clearly observe a trend of increasing performance over all algorithms when we decrease the proportion $\individual$ of users' individual feature.
However, for ConUCB, the benefit brought from smaller $\individual$ is limited.
The main reason is that absolute feedback is contaminated by user biases, which makes it difficult to share.
On the contrary, the improvement of \textbf{\updatediff} over other algorithms gets larger as $\individual$ gets smaller.
The trend validates our intuition that, compared to absolute feedback, relative feedback can better capture the intrinsic similarity by eliminating user biases.
This allows the comparison-based algorithms to further utilize feedback information.

\section{Related Work}\label{sec:related_work}
There are mainly two lines of research that are related to and motivate our work: conversational recommender system and preference-based learning.

\subsection{Conversational Recommender System} 
The most related works are a series of studies on conversational recommender systems.
Although traditional recommender systems have shown significant success in practice \cite{pazzani2007content, das2007google, mnih2007probabilistic}, there are still several shortages. 
One of the shortages is that the user's preference estimated from the historical interaction data by the traditional 
recommender system is static and thus the dynamic changes of user preference can not be well captured in this manner.
As an emerging topic in the community of recommender system, the conversational recommender system (CRS) is an appealing solution which aims to tackle the cold-start problem as well as the dynamic changes of user preferences \cite{gao2021advances,lei2020conversational}. 

Due to the significant developments of deep learning, many CRS equipped with natural language processing
techniques have been emerging in recent years \cite{fu2020tutorial}. 
\citet{li2018towards} propose a hierarchical recurrent encoder-decoder to address the cold-start problem in conjunction with classification of sentiment of dialogues.
\citet{yu2019visual} devise a visual dialog augmented model utilizing the user feedback presented in natural language.
\citet{zhou2020improving} adopt knowledge graphs to eliminate the semantic gap between natural language expression and item-level user preferences for more accurate recommendations. \citet{lei2020interactive} make conversations on graphs to enhance the model's explainability. 
Other works \cite{chen2019towards,zhang2018towards} utilize natural language understanding and generation to improve the accuracy of recommendations.

Some other works adopt reinforcement learning (RL) methods to address when and what to ask as the agent interacts with the users.
\citet{sun2018conversational} incorporate a deep policy network into the CRS to decide whether to conduct conversations or to recommend.
\citet{zhang2019text} design a constraint-augmented reinforcement learning framework to avoid recommendations violating user historical preferences.
\citet{zhou2020conversational} integrate the CRS, knowledge graph 
and bandit-based algorithm in order to face the fast-changing preferences and rich user expressions in music recommendation.
\citet{christakopoulou2016towards} first introduce the multi-armed bandits into CRS. 
And one of our main baselines, ConUCB \cite{zhang2020conversational} gives a clear and theoretically supported integration of linear contextual bandits and CRS.
A follow-up \cite{li2020seamlessly} proposes another bandit-based method for CRS using Thompson Sampling.


\subsection{Preference-Based Learning} 
Another line that is closely related to our work is preference-based learning \cite{wirth2017survey}.
In many RL scenarios such as games \cite{christiano2017deep} and robotics \cite{tucker2020preference},
the success of applications is highly dependent on a well-designed reward function.
However, users usually have difficulties in providing their absolute feedback explicitly, which makes it hard to define such a reward function.
In the meanwhile, the comparative preferences could better capture the user's intentions.
Therefore, a rich body of methods has been proposed to directly learn from users' preference.

\citet{christiano2017deep} use comparison feedback provided by humans rather than absolute numerical scores to train an agent without a well-defined reward function in a complex environment. 
In the area of robotics, \citet{tucker2020preference} and \citet{sadigh2017active} also use users' pairwise 
preferences for trajectories rather than 
numerical feedback to instruct the learning of robots.
In the field of information retrieval, \citet{joachims2017accurately} 
and \citet{radlinski2008does} find that user's clicks are 
sometimes biased and conclude that paired comparisons can be
more accurate than absolute numerical scores to evaluate retrieval quality.
To address the exploration-exploitation dilemma with constrained relative feedback, dueling bandits \cite{sui2018advancements} have been proposed and well-studied.
\citet{saha2019combinatorial} further explore the topic of combinatorial bandits with relative feedback provided instead of absolute bandit feedback.
\section{Conclusion}\label{sec:conclusion}
In this paper, we introduce a comparison-based conversation framework for item recommendation, utilizing relative feedback to better understand user preferences.
Within the framework, we develop the {\myalg} algorithm to effectively collect and incorporate relative bandit feedback.
To further explore the advantages of relative feedback, we propose to share relative feedback among similar users.
Extensive experiments on both synthetic and real-world datasets demonstrate the competitiveness of {\myalg}.
With relative feedback, our recommender system can efficiently elicit user preferences, especially among similar users, and provide satisfactory recommendations.

In the future, we plan to explore the following directions: 1) to improve the applicability of our framework, one may consider a more flexible system that can seamlessly query relative questions and absolute questions; or 2) one may focus on the feedback sharing mechanism to make it more robust and adaptive by incorporating techniques such as online clustering.



\bibliographystyle{ACM-Reference-Format}
\bibliography{normed_ref}


\begin{thebibliography}{32}


\ifx \showCODEN    \undefined \def \showCODEN     #1{\unskip}     \fi
\ifx \showDOI      \undefined \def \showDOI       #1{#1}\fi
\ifx \showISBNx    \undefined \def \showISBNx     #1{\unskip}     \fi
\ifx \showISBNxiii \undefined \def \showISBNxiii  #1{\unskip}     \fi
\ifx \showISSN     \undefined \def \showISSN      #1{\unskip}     \fi
\ifx \showLCCN     \undefined \def \showLCCN      #1{\unskip}     \fi
\ifx \shownote     \undefined \def \shownote      #1{#1}          \fi
\ifx \showarticletitle \undefined \def \showarticletitle #1{#1}   \fi
\ifx \showURL      \undefined \def \showURL       {\relax}        \fi
\providecommand\bibfield[2]{#2}
\providecommand\bibinfo[2]{#2}
\providecommand\natexlab[1]{#1}
\providecommand\showeprint[2][]{arXiv:#2}

\bibitem[\protect\citeauthoryear{Cantador, Brusilovsky, and Kuflik}{Cantador
  et~al\mbox{.}}{2011}]%
        {cantador2011second}
\bibfield{author}{\bibinfo{person}{Iv{\'{a}}n Cantador}, \bibinfo{person}{Peter
  Brusilovsky}, {and} \bibinfo{person}{Tsvi Kuflik}.}
  \bibinfo{year}{2011}\natexlab{}.
\newblock \showarticletitle{Second workshop on information heterogeneity and
  fusion in recommender systems (HetRec2011)}. In
  \bibinfo{booktitle}{\emph{Proceedings of the 2011 {ACM} Conference on
  Recommender Systems, RecSys 2011, Chicago, IL, USA, October 23-27, 2011}},
  \bibfield{editor}{\bibinfo{person}{Bamshad Mobasher},
  \bibinfo{person}{Robin~D. Burke}, \bibinfo{person}{Dietmar Jannach}, {and}
  \bibinfo{person}{Gediminas Adomavicius}} (Eds.). \bibinfo{publisher}{{ACM}},
  \bibinfo{pages}{387--388}.
\newblock
\urldef\tempurl%
\url{https://doi.org/10.1145/2043932.2044016}
\showDOI{\tempurl}


\bibitem[\protect\citeauthoryear{Chapelle, Joachims, Radlinski, and
  Yue}{Chapelle et~al\mbox{.}}{2012}]%
        {chapelle2012large}
\bibfield{author}{\bibinfo{person}{Olivier Chapelle}, \bibinfo{person}{Thorsten
  Joachims}, \bibinfo{person}{Filip Radlinski}, {and} \bibinfo{person}{Yisong
  Yue}.} \bibinfo{year}{2012}\natexlab{}.
\newblock \showarticletitle{Large-scale validation and analysis of interleaved
  search evaluation}.
\newblock \bibinfo{journal}{\emph{ACM Transactions on Information Systems
  (TOIS)}} \bibinfo{volume}{30}, \bibinfo{number}{1} (\bibinfo{year}{2012}),
  \bibinfo{pages}{1--41}.
\newblock


\bibitem[\protect\citeauthoryear{Chen, Lin, Zhang, Ding, Cen, Yang, and
  Tang}{Chen et~al\mbox{.}}{2019}]%
        {chen2019towards}
\bibfield{author}{\bibinfo{person}{Qibin Chen}, \bibinfo{person}{Junyang Lin},
  \bibinfo{person}{Yichang Zhang}, \bibinfo{person}{Ming Ding},
  \bibinfo{person}{Yukuo Cen}, \bibinfo{person}{Hongxia Yang}, {and}
  \bibinfo{person}{Jie Tang}.} \bibinfo{year}{2019}\natexlab{}.
\newblock \showarticletitle{Towards Knowledge-Based Recommender Dialog System}.
  In \bibinfo{booktitle}{\emph{Proceedings of the 2019 Conference on Empirical
  Methods in Natural Language Processing and the 9th International Joint
  Conference on Natural Language Processing (EMNLP-IJCNLP)}}.
  \bibinfo{publisher}{Association for Computational Linguistics},
  \bibinfo{address}{Hong Kong, China}, \bibinfo{pages}{1803--1813}.
\newblock
\urldef\tempurl%
\url{https://doi.org/10.18653/v1/D19-1189}
\showDOI{\tempurl}


\bibitem[\protect\citeauthoryear{Chin, Yuan, Yang, Zhuang, Juan, and Lin}{Chin
  et~al\mbox{.}}{2016}]%
        {chin2016libmf}
\bibfield{author}{\bibinfo{person}{Wei-Sheng Chin}, \bibinfo{person}{Bo-Wen
  Yuan}, \bibinfo{person}{Meng-Yuan Yang}, \bibinfo{person}{Yong Zhuang},
  \bibinfo{person}{Yu-Chin Juan}, {and} \bibinfo{person}{Chih-Jen Lin}.}
  \bibinfo{year}{2016}\natexlab{}.
\newblock \showarticletitle{LIBMF: A Library for Parallel Matrix Factorization
  in Shared-memory Systems}.
\newblock \bibinfo{journal}{\emph{Journal of Machine Learning Research}}
  \bibinfo{volume}{17}, \bibinfo{number}{86} (\bibinfo{year}{2016}),
  \bibinfo{pages}{1--5}.
\newblock
\urldef\tempurl%
\url{http://jmlr.org/papers/v17/15-471.html}
\showURL{%
\tempurl}


\bibitem[\protect\citeauthoryear{Christakopoulou, Radlinski, and
  Hofmann}{Christakopoulou et~al\mbox{.}}{2016}]%
        {christakopoulou2016towards}
\bibfield{author}{\bibinfo{person}{Konstantina Christakopoulou},
  \bibinfo{person}{Filip Radlinski}, {and} \bibinfo{person}{Katja Hofmann}.}
  \bibinfo{year}{2016}\natexlab{}.
\newblock \showarticletitle{Towards Conversational Recommender Systems}. In
  \bibinfo{booktitle}{\emph{Proceedings of the 22nd {ACM} {SIGKDD}
  International Conference on Knowledge Discovery and Data Mining, San
  Francisco, CA, USA, August 13-17, 2016}},
  \bibfield{editor}{\bibinfo{person}{Balaji Krishnapuram},
  \bibinfo{person}{Mohak Shah}, \bibinfo{person}{Alexander~J. Smola},
  \bibinfo{person}{Charu~C. Aggarwal}, \bibinfo{person}{Dou Shen}, {and}
  \bibinfo{person}{Rajeev Rastogi}} (Eds.). \bibinfo{publisher}{{ACM}},
  \bibinfo{pages}{815--824}.
\newblock
\urldef\tempurl%
\url{https://doi.org/10.1145/2939672.2939746}
\showDOI{\tempurl}


\bibitem[\protect\citeauthoryear{Christiano, Leike, Brown, Martic, Legg, and
  Amodei}{Christiano et~al\mbox{.}}{2017}]%
        {christiano2017deep}
\bibfield{author}{\bibinfo{person}{Paul~F. Christiano}, \bibinfo{person}{Jan
  Leike}, \bibinfo{person}{Tom~B. Brown}, \bibinfo{person}{Miljan Martic},
  \bibinfo{person}{Shane Legg}, {and} \bibinfo{person}{Dario Amodei}.}
  \bibinfo{year}{2017}\natexlab{}.
\newblock \showarticletitle{Deep Reinforcement Learning from Human
  Preferences}. In \bibinfo{booktitle}{\emph{Advances in Neural Information
  Processing Systems 30: Annual Conference on Neural Information Processing
  Systems 2017, December 4-9, 2017, Long Beach, CA, {USA}}},
  \bibfield{editor}{\bibinfo{person}{Isabelle Guyon}, \bibinfo{person}{Ulrike
  von Luxburg}, \bibinfo{person}{Samy Bengio}, \bibinfo{person}{Hanna~M.
  Wallach}, \bibinfo{person}{Rob Fergus}, \bibinfo{person}{S.~V.~N.
  Vishwanathan}, {and} \bibinfo{person}{Roman Garnett}} (Eds.).
  \bibinfo{pages}{4299--4307}.
\newblock
\urldef\tempurl%
\url{https://proceedings.neurips.cc/paper/2017/hash/d5e2c0adad503c91f91df240d0cd4e49-Abstract.html}
\showURL{%
\tempurl}


\bibitem[\protect\citeauthoryear{Das, Datar, Garg, and Rajaram}{Das
  et~al\mbox{.}}{2007}]%
        {das2007google}
\bibfield{author}{\bibinfo{person}{Abhinandan Das}, \bibinfo{person}{Mayur
  Datar}, \bibinfo{person}{Ashutosh Garg}, {and} \bibinfo{person}{Shyamsundar
  Rajaram}.} \bibinfo{year}{2007}\natexlab{}.
\newblock \showarticletitle{Google news personalization: scalable online
  collaborative filtering}. In \bibinfo{booktitle}{\emph{Proceedings of the
  16th International Conference on World Wide Web, {WWW} 2007, Banff, Alberta,
  Canada, May 8-12, 2007}}, \bibfield{editor}{\bibinfo{person}{Carey~L.
  Williamson}, \bibinfo{person}{Mary~Ellen Zurko}, \bibinfo{person}{Peter~F.
  Patel{-}Schneider}, {and} \bibinfo{person}{Prashant~J. Shenoy}} (Eds.).
  \bibinfo{publisher}{{ACM}}, \bibinfo{pages}{271--280}.
\newblock
\urldef\tempurl%
\url{https://doi.org/10.1145/1242572.1242610}
\showDOI{\tempurl}


\bibitem[\protect\citeauthoryear{Fu, Xian, Zhang, and Zhang}{Fu
  et~al\mbox{.}}{2020}]%
        {fu2020tutorial}
\bibfield{author}{\bibinfo{person}{Zuohui Fu}, \bibinfo{person}{Yikun Xian},
  \bibinfo{person}{Yongfeng Zhang}, {and} \bibinfo{person}{Yi Zhang}.}
  \bibinfo{year}{2020}\natexlab{}.
\newblock \showarticletitle{Tutorial on Conversational Recommendation Systems}.
  In \bibinfo{booktitle}{\emph{RecSys 2020: Fourteenth {ACM} Conference on
  Recommender Systems, Virtual Event, Brazil, September 22-26, 2020}},
  \bibfield{editor}{\bibinfo{person}{Rodrygo L.~T. Santos},
  \bibinfo{person}{Leandro~Balby Marinho}, \bibinfo{person}{Elizabeth~M. Daly},
  \bibinfo{person}{Li~Chen}, \bibinfo{person}{Kim Falk}, \bibinfo{person}{Noam
  Koenigstein}, {and} \bibinfo{person}{Edleno~Silva de~Moura}} (Eds.).
  \bibinfo{publisher}{{ACM}}, \bibinfo{pages}{751--753}.
\newblock
\urldef\tempurl%
\url{https://doi.org/10.1145/3383313.3411548}
\showDOI{\tempurl}


\bibitem[\protect\citeauthoryear{Gao, Lei, He, de~Rijke, and Chua}{Gao
  et~al\mbox{.}}{2021}]%
        {gao2021advances}
\bibfield{author}{\bibinfo{person}{Chongming Gao}, \bibinfo{person}{Wenqiang
  Lei}, \bibinfo{person}{Xiangnan He}, \bibinfo{person}{Maarten de Rijke},
  {and} \bibinfo{person}{Tat{-}Seng Chua}.} \bibinfo{year}{2021}\natexlab{}.
\newblock \showarticletitle{Advances and Challenges in Conversational
  Recommender Systems: {A} Survey}.
\newblock \bibinfo{journal}{\emph{CoRR}}  \bibinfo{volume}{abs/2101.09459}
  (\bibinfo{year}{2021}).
\newblock
\showeprint[arxiv]{2101.09459}
\urldef\tempurl%
\url{https://arxiv.org/abs/2101.09459}
\showURL{%
\tempurl}


\bibitem[\protect\citeauthoryear{Joachims, Granka, Pan, Hembrooke, and
  Gay}{Joachims et~al\mbox{.}}{2017}]%
        {joachims2017accurately}
\bibfield{author}{\bibinfo{person}{Thorsten Joachims}, \bibinfo{person}{Laura
  Granka}, \bibinfo{person}{Bing Pan}, \bibinfo{person}{Helene Hembrooke},
  {and} \bibinfo{person}{Geri Gay}.} \bibinfo{year}{2017}\natexlab{}.
\newblock \showarticletitle{Accurately interpreting clickthrough data as
  implicit feedback}. In \bibinfo{booktitle}{\emph{ACM SIGIR Forum}},
  Vol.~\bibinfo{volume}{51}. Acm New York, NY, USA, \bibinfo{pages}{4--11}.
\newblock


\bibitem[\protect\citeauthoryear{Lei, He, de~Rijke, and Chua}{Lei
  et~al\mbox{.}}{2020a}]%
        {lei2020conversational}
\bibfield{author}{\bibinfo{person}{Wenqiang Lei}, \bibinfo{person}{Xiangnan
  He}, \bibinfo{person}{Maarten de Rijke}, {and} \bibinfo{person}{Tat{-}Seng
  Chua}.} \bibinfo{year}{2020}\natexlab{a}.
\newblock \showarticletitle{Conversational Recommendation: Formulation,
  Methods, and Evaluation}. In \bibinfo{booktitle}{\emph{Proceedings of the
  43rd International {ACM} {SIGIR} conference on research and development in
  Information Retrieval, {SIGIR} 2020, Virtual Event, China, July 25-30,
  2020}}, \bibfield{editor}{\bibinfo{person}{Jimmy Huang},
  \bibinfo{person}{Yi~Chang}, \bibinfo{person}{Xueqi Cheng},
  \bibinfo{person}{Jaap Kamps}, \bibinfo{person}{Vanessa Murdock},
  \bibinfo{person}{Ji{-}Rong Wen}, {and} \bibinfo{person}{Yiqun Liu}} (Eds.).
  \bibinfo{publisher}{{ACM}}, \bibinfo{pages}{2425--2428}.
\newblock
\urldef\tempurl%
\url{https://doi.org/10.1145/3397271.3401419}
\showDOI{\tempurl}


\bibitem[\protect\citeauthoryear{Lei, Zhang, He, Miao, Wang, Chen, and
  Chua}{Lei et~al\mbox{.}}{2020b}]%
        {lei2020interactive}
\bibfield{author}{\bibinfo{person}{Wenqiang Lei}, \bibinfo{person}{Gangyi
  Zhang}, \bibinfo{person}{Xiangnan He}, \bibinfo{person}{Yisong Miao},
  \bibinfo{person}{Xiang Wang}, \bibinfo{person}{Liang Chen}, {and}
  \bibinfo{person}{Tat{-}Seng Chua}.} \bibinfo{year}{2020}\natexlab{b}.
\newblock \showarticletitle{Interactive Path Reasoning on Graph for
  Conversational Recommendation}. In \bibinfo{booktitle}{\emph{{KDD} '20: The
  26th {ACM} {SIGKDD} Conference on Knowledge Discovery and Data Mining,
  Virtual Event, CA, USA, August 23-27, 2020}},
  \bibfield{editor}{\bibinfo{person}{Rajesh Gupta}, \bibinfo{person}{Yan Liu},
  \bibinfo{person}{Jiliang Tang}, {and} \bibinfo{person}{B.~Aditya Prakash}}
  (Eds.). \bibinfo{publisher}{{ACM}}, \bibinfo{pages}{2073--2083}.
\newblock
\urldef\tempurl%
\url{https://dl.acm.org/doi/10.1145/3394486.3403258}
\showURL{%
\tempurl}


\bibitem[\protect\citeauthoryear{Li, Chu, Langford, and Schapire}{Li
  et~al\mbox{.}}{2010}]%
        {li2010contextual}
\bibfield{author}{\bibinfo{person}{Lihong Li}, \bibinfo{person}{Wei Chu},
  \bibinfo{person}{John Langford}, {and} \bibinfo{person}{Robert~E. Schapire}.}
  \bibinfo{year}{2010}\natexlab{}.
\newblock \showarticletitle{A contextual-bandit approach to personalized news
  article recommendation}. In \bibinfo{booktitle}{\emph{Proceedings of the 19th
  International Conference on World Wide Web, {WWW} 2010, Raleigh, North
  Carolina, USA, April 26-30, 2010}},
  \bibfield{editor}{\bibinfo{person}{Michael Rappa}, \bibinfo{person}{Paul
  Jones}, \bibinfo{person}{Juliana Freire}, {and} \bibinfo{person}{Soumen
  Chakrabarti}} (Eds.). \bibinfo{publisher}{{ACM}}, \bibinfo{pages}{661--670}.
\newblock
\urldef\tempurl%
\url{https://doi.org/10.1145/1772690.1772758}
\showDOI{\tempurl}


\bibitem[\protect\citeauthoryear{Li, Kahou, Schulz, Michalski, Charlin, and
  Pal}{Li et~al\mbox{.}}{2018}]%
        {li2018towards}
\bibfield{author}{\bibinfo{person}{Raymond Li},
  \bibinfo{person}{Samira~Ebrahimi Kahou}, \bibinfo{person}{Hannes Schulz},
  \bibinfo{person}{Vincent Michalski}, \bibinfo{person}{Laurent Charlin}, {and}
  \bibinfo{person}{Chris Pal}.} \bibinfo{year}{2018}\natexlab{}.
\newblock \showarticletitle{Towards Deep Conversational Recommendations}. In
  \bibinfo{booktitle}{\emph{Advances in Neural Information Processing Systems
  31: Annual Conference on Neural Information Processing Systems 2018, NeurIPS
  2018, December 3-8, 2018, Montr{\'{e}}al, Canada}},
  \bibfield{editor}{\bibinfo{person}{Samy Bengio}, \bibinfo{person}{Hanna~M.
  Wallach}, \bibinfo{person}{Hugo Larochelle}, \bibinfo{person}{Kristen
  Grauman}, \bibinfo{person}{Nicol{\`{o}} Cesa{-}Bianchi}, {and}
  \bibinfo{person}{Roman Garnett}} (Eds.). \bibinfo{pages}{9748--9758}.
\newblock
\urldef\tempurl%
\url{https://proceedings.neurips.cc/paper/2018/hash/800de15c79c8d840f4e78d3af937d4d4-Abstract.html}
\showURL{%
\tempurl}


\bibitem[\protect\citeauthoryear{Li, Lei, Wu, He, Jiang, and Chua}{Li
  et~al\mbox{.}}{2020}]%
        {li2020seamlessly}
\bibfield{author}{\bibinfo{person}{Shijun Li}, \bibinfo{person}{Wenqiang Lei},
  \bibinfo{person}{Qingyun Wu}, \bibinfo{person}{Xiangnan He},
  \bibinfo{person}{Peng Jiang}, {and} \bibinfo{person}{Tat{-}Seng Chua}.}
  \bibinfo{year}{2020}\natexlab{}.
\newblock \showarticletitle{Seamlessly Unifying Attributes and Items:
  Conversational Recommendation for Cold-Start Users}.
\newblock \bibinfo{journal}{\emph{CoRR}}  \bibinfo{volume}{abs/2005.12979}
  (\bibinfo{year}{2020}).
\newblock
\showeprint[arxiv]{2005.12979}
\urldef\tempurl%
\url{https://arxiv.org/abs/2005.12979}
\showURL{%
\tempurl}


\bibitem[\protect\citeauthoryear{Pazzani and Billsus}{Pazzani and
  Billsus}{2007}]%
        {pazzani2007content}
\bibfield{author}{\bibinfo{person}{Michael~J Pazzani} {and}
  \bibinfo{person}{Daniel Billsus}.} \bibinfo{year}{2007}\natexlab{}.
\newblock \showarticletitle{Content-based recommendation systems}.
\newblock In \bibinfo{booktitle}{\emph{The adaptive web}}.
  \bibinfo{publisher}{Springer}, \bibinfo{pages}{325--341}.
\newblock


\bibitem[\protect\citeauthoryear{Radlinski, Kurup, and Joachims}{Radlinski
  et~al\mbox{.}}{2008}]%
        {radlinski2008does}
\bibfield{author}{\bibinfo{person}{Filip Radlinski}, \bibinfo{person}{Madhu
  Kurup}, {and} \bibinfo{person}{Thorsten Joachims}.}
  \bibinfo{year}{2008}\natexlab{}.
\newblock \showarticletitle{How does clickthrough data reflect retrieval
  quality?}. In \bibinfo{booktitle}{\emph{Proceedings of the 17th ACM
  conference on Information and knowledge management}}.
  \bibinfo{pages}{43--52}.
\newblock


\bibitem[\protect\citeauthoryear{Rumelhart, Hinton, and Williams}{Rumelhart
  et~al\mbox{.}}{1985}]%
        {rumelhart1985learning}
\bibfield{author}{\bibinfo{person}{David~E Rumelhart},
  \bibinfo{person}{Geoffrey~E Hinton}, {and} \bibinfo{person}{Ronald~J
  Williams}.} \bibinfo{year}{1985}\natexlab{}.
\newblock \bibinfo{booktitle}{\emph{Learning internal representations by error
  propagation}}.
\newblock \bibinfo{type}{{T}echnical {R}eport}.
  \bibinfo{institution}{California Univ San Diego La Jolla Inst for Cognitive
  Science}.
\newblock


\bibitem[\protect\citeauthoryear{Sadigh, Dragan, Sastry, and Seshia}{Sadigh
  et~al\mbox{.}}{2017}]%
        {sadigh2017active}
\bibfield{author}{\bibinfo{person}{Dorsa Sadigh}, \bibinfo{person}{Anca~D
  Dragan}, \bibinfo{person}{Shankar Sastry}, {and} \bibinfo{person}{Sanjit~A
  Seshia}.} \bibinfo{year}{2017}\natexlab{}.
\newblock \showarticletitle{Active Preference-Based Learning of Reward
  Functions.}. In \bibinfo{booktitle}{\emph{Robotics: Science and Systems}}.
\newblock


\bibitem[\protect\citeauthoryear{Saha and Gopalan}{Saha and Gopalan}{2019}]%
        {saha2019combinatorial}
\bibfield{author}{\bibinfo{person}{Aadirupa Saha} {and} \bibinfo{person}{Aditya
  Gopalan}.} \bibinfo{year}{2019}\natexlab{}.
\newblock \showarticletitle{Combinatorial Bandits with Relative Feedback}. In
  \bibinfo{booktitle}{\emph{Advances in Neural Information Processing Systems
  32: Annual Conference on Neural Information Processing Systems 2019, NeurIPS
  2019, December 8-14, 2019, Vancouver, BC, Canada}},
  \bibfield{editor}{\bibinfo{person}{Hanna~M. Wallach}, \bibinfo{person}{Hugo
  Larochelle}, \bibinfo{person}{Alina Beygelzimer}, \bibinfo{person}{Florence
  d'Alch{\'{e}}{-}Buc}, \bibinfo{person}{Emily~B. Fox}, {and}
  \bibinfo{person}{Roman Garnett}} (Eds.). \bibinfo{pages}{983--993}.
\newblock
\urldef\tempurl%
\url{https://proceedings.neurips.cc/paper/2019/hash/5e388103a391daabe3de1d76a6739ccd-Abstract.html}
\showURL{%
\tempurl}


\bibitem[\protect\citeauthoryear{Salakhutdinov and Mnih}{Salakhutdinov and
  Mnih}{2007}]%
        {mnih2007probabilistic}
\bibfield{author}{\bibinfo{person}{Ruslan Salakhutdinov} {and}
  \bibinfo{person}{Andriy Mnih}.} \bibinfo{year}{2007}\natexlab{}.
\newblock \showarticletitle{Probabilistic Matrix Factorization}. In
  \bibinfo{booktitle}{\emph{Advances in Neural Information Processing Systems
  20, Proceedings of the Twenty-First Annual Conference on Neural Information
  Processing Systems, Vancouver, British Columbia, Canada, December 3-6,
  2007}}, \bibfield{editor}{\bibinfo{person}{John~C. Platt},
  \bibinfo{person}{Daphne Koller}, \bibinfo{person}{Yoram Singer}, {and}
  \bibinfo{person}{Sam~T. Roweis}} (Eds.). \bibinfo{publisher}{Curran
  Associates, Inc.}, \bibinfo{pages}{1257--1264}.
\newblock
\urldef\tempurl%
\url{https://proceedings.neurips.cc/paper/2007/hash/d7322ed717dedf1eb4e6e52a37ea7bcd-Abstract.html}
\showURL{%
\tempurl}


\bibitem[\protect\citeauthoryear{Sui, Zoghi, Hofmann, and Yue}{Sui
  et~al\mbox{.}}{2018}]%
        {sui2018advancements}
\bibfield{author}{\bibinfo{person}{Yanan Sui}, \bibinfo{person}{Masrour Zoghi},
  \bibinfo{person}{Katja Hofmann}, {and} \bibinfo{person}{Yisong Yue}.}
  \bibinfo{year}{2018}\natexlab{}.
\newblock \showarticletitle{Advancements in Dueling Bandits}. In
  \bibinfo{booktitle}{\emph{Proceedings of the Twenty-Seventh International
  Joint Conference on Artificial Intelligence, {IJCAI} 2018, July 13-19, 2018,
  Stockholm, Sweden}}, \bibfield{editor}{\bibinfo{person}{J{\'{e}}r{\^{o}}me
  Lang}} (Ed.). \bibinfo{publisher}{ijcai.org}, \bibinfo{pages}{5502--5510}.
\newblock
\urldef\tempurl%
\url{https://doi.org/10.24963/ijcai.2018/776}
\showDOI{\tempurl}


\bibitem[\protect\citeauthoryear{Sun and Zhang}{Sun and Zhang}{2018}]%
        {sun2018conversational}
\bibfield{author}{\bibinfo{person}{Yueming Sun} {and} \bibinfo{person}{Yi
  Zhang}.} \bibinfo{year}{2018}\natexlab{}.
\newblock \showarticletitle{Conversational Recommender System}. In
  \bibinfo{booktitle}{\emph{The 41st International {ACM} {SIGIR} Conference on
  Research {\&} Development in Information Retrieval, {SIGIR} 2018, Ann Arbor,
  MI, USA, July 08-12, 2018}}, \bibfield{editor}{\bibinfo{person}{Kevyn
  Collins{-}Thompson}, \bibinfo{person}{Qiaozhu Mei}, \bibinfo{person}{Brian~D.
  Davison}, \bibinfo{person}{Yiqun Liu}, {and} \bibinfo{person}{Emine Yilmaz}}
  (Eds.). \bibinfo{publisher}{{ACM}}, \bibinfo{pages}{235--244}.
\newblock
\urldef\tempurl%
\url{https://doi.org/10.1145/3209978.3210002}
\showDOI{\tempurl}


\bibitem[\protect\citeauthoryear{Tucker, Novoseller, Kann, Sui, Yue, Burdick,
  and Ames}{Tucker et~al\mbox{.}}{2020}]%
        {tucker2020preference}
\bibfield{author}{\bibinfo{person}{Maegan Tucker}, \bibinfo{person}{Ellen
  Novoseller}, \bibinfo{person}{Claudia Kann}, \bibinfo{person}{Yanan Sui},
  \bibinfo{person}{Yisong Yue}, \bibinfo{person}{Joel~W Burdick}, {and}
  \bibinfo{person}{Aaron~D Ames}.} \bibinfo{year}{2020}\natexlab{}.
\newblock \showarticletitle{Preference-based learning for exoskeleton gait
  optimization}. In \bibinfo{booktitle}{\emph{2020 IEEE International
  Conference on Robotics and Automation (ICRA)}}. IEEE,
  \bibinfo{pages}{2351--2357}.
\newblock


\bibitem[\protect\citeauthoryear{Wirth, Akrour, Neumann, F{\"u}rnkranz,
  et~al\mbox{.}}{Wirth et~al\mbox{.}}{2017}]%
        {wirth2017survey}
\bibfield{author}{\bibinfo{person}{Christian Wirth}, \bibinfo{person}{Riad
  Akrour}, \bibinfo{person}{Gerhard Neumann}, \bibinfo{person}{Johannes
  F{\"u}rnkranz}, {et~al\mbox{.}}} \bibinfo{year}{2017}\natexlab{}.
\newblock \showarticletitle{A survey of preference-based reinforcement learning
  methods}.
\newblock \bibinfo{journal}{\emph{Journal of Machine Learning Research}}
  \bibinfo{volume}{18}, \bibinfo{number}{136} (\bibinfo{year}{2017}),
  \bibinfo{pages}{1--46}.
\newblock


\bibitem[\protect\citeauthoryear{Yu, Shen, and Jin}{Yu et~al\mbox{.}}{2019}]%
        {yu2019visual}
\bibfield{author}{\bibinfo{person}{Tong Yu}, \bibinfo{person}{Yilin Shen},
  {and} \bibinfo{person}{Hongxia Jin}.} \bibinfo{year}{2019}\natexlab{}.
\newblock \showarticletitle{A Visual Dialog Augmented Interactive Recommender
  System}. In \bibinfo{booktitle}{\emph{Proceedings of the 25th {ACM} {SIGKDD}
  International Conference on Knowledge Discovery {\&} Data Mining, {KDD} 2019,
  Anchorage, AK, USA, August 4-8, 2019}},
  \bibfield{editor}{\bibinfo{person}{Ankur Teredesai}, \bibinfo{person}{Vipin
  Kumar}, \bibinfo{person}{Ying Li}, \bibinfo{person}{R{\'{o}}mer Rosales},
  \bibinfo{person}{Evimaria Terzi}, {and} \bibinfo{person}{George Karypis}}
  (Eds.). \bibinfo{publisher}{{ACM}}, \bibinfo{pages}{157--165}.
\newblock
\urldef\tempurl%
\url{https://doi.org/10.1145/3292500.3330991}
\showDOI{\tempurl}


\bibitem[\protect\citeauthoryear{Yue and Joachims}{Yue and Joachims}{2009}]%
        {yue2009interactively}
\bibfield{author}{\bibinfo{person}{Yisong Yue} {and} \bibinfo{person}{Thorsten
  Joachims}.} \bibinfo{year}{2009}\natexlab{}.
\newblock \showarticletitle{Interactively optimizing information retrieval
  systems as a dueling bandits problem}. In
  \bibinfo{booktitle}{\emph{Proceedings of the 26th Annual International
  Conference on Machine Learning, {ICML} 2009, Montreal, Quebec, Canada, June
  14-18, 2009}} \emph{(\bibinfo{series}{{ACM} International Conference
  Proceeding Series}, Vol.~\bibinfo{volume}{382})},
  \bibfield{editor}{\bibinfo{person}{Andrea~Pohoreckyj Danyluk},
  \bibinfo{person}{L{\'{e}}on Bottou}, {and} \bibinfo{person}{Michael~L.
  Littman}} (Eds.). \bibinfo{publisher}{{ACM}}, \bibinfo{pages}{1201--1208}.
\newblock
\urldef\tempurl%
\url{https://doi.org/10.1145/1553374.1553527}
\showDOI{\tempurl}


\bibitem[\protect\citeauthoryear{Zhang, Yu, Shen, Jin, Chen, and Carin}{Zhang
  et~al\mbox{.}}{2019}]%
        {zhang2019text}
\bibfield{author}{\bibinfo{person}{Ruiyi Zhang}, \bibinfo{person}{Tong Yu},
  \bibinfo{person}{Yilin Shen}, \bibinfo{person}{Hongxia Jin},
  \bibinfo{person}{Changyou Chen}, {and} \bibinfo{person}{Lawrence Carin}.}
  \bibinfo{year}{2019}\natexlab{}.
\newblock \showarticletitle{Text-Based Interactive Recommendation with
  Constraint-Augmented Reinforcement Learning}.
\newblock  (\bibinfo{year}{2019}).
\newblock


\bibitem[\protect\citeauthoryear{Zhang, Xie, Li, and Lui}{Zhang
  et~al\mbox{.}}{2020}]%
        {zhang2020conversational}
\bibfield{author}{\bibinfo{person}{Xiaoying Zhang}, \bibinfo{person}{Hong Xie},
  \bibinfo{person}{Hang Li}, {and} \bibinfo{person}{John C.~S. Lui}.}
  \bibinfo{year}{2020}\natexlab{}.
\newblock \showarticletitle{Conversational Contextual Bandit: Algorithm and
  Application}. In \bibinfo{booktitle}{\emph{{WWW} '20: The Web Conference
  2020, Taipei, Taiwan, April 20-24, 2020}},
  \bibfield{editor}{\bibinfo{person}{Yennun Huang}, \bibinfo{person}{Irwin
  King}, \bibinfo{person}{Tie{-}Yan Liu}, {and} \bibinfo{person}{Maarten van
  Steen}} (Eds.). \bibinfo{publisher}{{ACM} / {IW3C2}},
  \bibinfo{pages}{662--672}.
\newblock
\urldef\tempurl%
\url{https://doi.org/10.1145/3366423.3380148}
\showDOI{\tempurl}


\bibitem[\protect\citeauthoryear{Zhang, Chen, Ai, Yang, and Croft}{Zhang
  et~al\mbox{.}}{2018}]%
        {zhang2018towards}
\bibfield{author}{\bibinfo{person}{Yongfeng Zhang}, \bibinfo{person}{Xu Chen},
  \bibinfo{person}{Qingyao Ai}, \bibinfo{person}{Liu Yang}, {and}
  \bibinfo{person}{W.~Bruce Croft}.} \bibinfo{year}{2018}\natexlab{}.
\newblock \showarticletitle{Towards Conversational Search and Recommendation:
  System Ask, User Respond}. In \bibinfo{booktitle}{\emph{Proceedings of the
  27th {ACM} International Conference on Information and Knowledge Management,
  {CIKM} 2018, Torino, Italy, October 22-26, 2018}},
  \bibfield{editor}{\bibinfo{person}{Alfredo Cuzzocrea}, \bibinfo{person}{James
  Allan}, \bibinfo{person}{Norman~W. Paton}, \bibinfo{person}{Divesh
  Srivastava}, \bibinfo{person}{Rakesh Agrawal}, \bibinfo{person}{Andrei~Z.
  Broder}, \bibinfo{person}{Mohammed~J. Zaki}, \bibinfo{person}{K.~Sel{\c{c}}uk
  Candan}, \bibinfo{person}{Alexandros Labrinidis}, \bibinfo{person}{Assaf
  Schuster}, {and} \bibinfo{person}{Haixun Wang}} (Eds.).
  \bibinfo{publisher}{{ACM}}, \bibinfo{pages}{177--186}.
\newblock
\urldef\tempurl%
\url{https://doi.org/10.1145/3269206.3271776}
\showDOI{\tempurl}


\bibitem[\protect\citeauthoryear{Zhou, Jin, Wang, and Zhang}{Zhou
  et~al\mbox{.}}{2020a}]%
        {zhou2020conversational}
\bibfield{author}{\bibinfo{person}{Chunyi Zhou}, \bibinfo{person}{Yuanyuan
  Jin}, \bibinfo{person}{Xiaoling Wang}, {and} \bibinfo{person}{Yingjie
  Zhang}.} \bibinfo{year}{2020}\natexlab{a}.
\newblock \showarticletitle{Conversational Music Recommendation based on
  Bandits}. In \bibinfo{booktitle}{\emph{2020 IEEE International Conference on
  Knowledge Graph (ICKG)}}. IEEE, \bibinfo{pages}{41--48}.
\newblock


\bibitem[\protect\citeauthoryear{Zhou, Zhao, Bian, Zhou, Wen, and Yu}{Zhou
  et~al\mbox{.}}{2020b}]%
        {zhou2020improving}
\bibfield{author}{\bibinfo{person}{Kun Zhou}, \bibinfo{person}{Wayne~Xin Zhao},
  \bibinfo{person}{Shuqing Bian}, \bibinfo{person}{Yuanhang Zhou},
  \bibinfo{person}{Ji{-}Rong Wen}, {and} \bibinfo{person}{Jingsong Yu}.}
  \bibinfo{year}{2020}\natexlab{b}.
\newblock \showarticletitle{Improving Conversational Recommender Systems via
  Knowledge Graph based Semantic Fusion}. In \bibinfo{booktitle}{\emph{{KDD}
  '20: The 26th {ACM} {SIGKDD} Conference on Knowledge Discovery and Data
  Mining, Virtual Event, CA, USA, August 23-27, 2020}},
  \bibfield{editor}{\bibinfo{person}{Rajesh Gupta}, \bibinfo{person}{Yan Liu},
  \bibinfo{person}{Jiliang Tang}, {and} \bibinfo{person}{B.~Aditya Prakash}}
  (Eds.). \bibinfo{publisher}{{ACM}}, \bibinfo{pages}{1006--1014}.
\newblock
\urldef\tempurl%
\url{https://dl.acm.org/doi/10.1145/3394486.3403143}
\showURL{%
\tempurl}


\end{thebibliography}

\appendix

\end{document}